\newcommand{\ndy}{(Nd$_x$Y$_{1-x}$)$_{2}$BaNiO$_{5}$}
\newcommand{\nd}{Nd$_{2}$BaNiO$_{5}$}
\newcommand{\pr}{Pr$_{2}$BaNiO$_{5}$}
\newcommand{\rr}{$R_{2}$BaNiO$_{5}$}
\newcommand{\nii}{Ni$^{2+}$}

\newcommand{\rion}{$R^{3+}$}

\documentstyle[prb,aps,preprint]{revtex}
\begin{document}
\draft
\title{The role of single-ion excitations in the mixed-spin quasi-1-dimensional quantum antiferromagnet \nd.}

\author{A. Zheludev and S. Maslov}
\address{Physics Department, Brookhaven National Laboratory, Upton,
New York 11973 USA}

\author{T. Yokoo and J. Akimitsu}
\address{Department of Physics, Aoyama-Gakuin University, 6-16-1,
Chitosedai, Setagaya-ku, Tokyo 157 Japan}

\author{ S. Raymond}
\address{Université de Geneve DPMC 24, quai Ernest Ansermet 1211, Geneve 4,
Switzerland.}

\author{S. E. Nagler}
\address{Oak Ridge National Laboratory, Bld. 7692, MS 6393, P.O. Box 2008, Oak Ridge, TN
37831, USA.}

\author{K. Hirota}
\address{Department of Physics, Tohoku
University, Sendai 980-8578, Japan CREST, JST, Japan.}

\date{\today}
\maketitle
\begin{abstract}
Inelastic neutron scattering experiments on \nd\ single-crystals
and powder samples are used to study the dynamic coupling of
1-dimensional Haldane-gap excitations in the $S=1$
Ni$^{2+}$-chains to local crystal-field transitions, associated
with the rare earth ions. Substantial interference between the two
types of excitations is observed even in the 1-dimensional
paramagnetic phase. Despite that, the results provide solid
justification for the previously proposed ``static staggered
field'' model for \rr nickelates. The observed behaviour is
qualitatively explained by a simple
chain-Random-Phase-Approximation (chain-RPA) model.
\end{abstract}


\narrowtext

\section{Introduction}
Linear-chain rare earth nickelates with the general formula \rr\
were recently recognized as unique model compounds for
experimental studies of Haldane-gap quantum spin chains in strong
{\it staggered} applied fields.\cite{Zheludev99NN} In these
materials the Haldane subsystem is formed by chains of $S=1$ \nii\
ions with relatively strong, almost isotropic, nearest-neighbor
antiferromagnetic (AF) exchange interactions ($J\approx 27$~meV).
The corresponding Haldane energy gap in the magnetic excitation
spectrum is $\Delta\approx
11$~meV.\cite{YBANO,Zheludev96PBANO,Zheludev96NBANO,Yokoo97NDY,Yokoo98,Raymond99}
An effective staggered exchange field $H_{\pi}$ acting on these
1-dimensional (1-D) objects is generated by 3-dimensional (3-D)
magnetic long range ordering in the ``auxiliary'' sublattice of
magnetic rare earth ions, that typically occurs at N\'{e}el
temperatures of several tens of Kelvin.\cite{Garcia} The magnitude
of the staggered field is proportional to the ordered moment on
the \rion\ sites, and can be varied in an experiment, by changing
the temperature. Static magnetic properties of \rr\ compounds are
rather well described by the chain-Mean Field (MF)
model.\cite{ZM98NBANO-L,Yokoo98} Using this theory to analyze the
measured temperature dependencies of magnetic order parameters for
a number of \rr\ species, previously enabled us to extract the
{\it staggered magnetization function} for a single Haldane spin
chain.\cite{ZM98NBANO-L} These experimental results were found to
be in excellent agreement with theoretical calculations for
isolated chains.\cite{MZ98-L,numerics} The view of \rr~systems as
Haldane spin chains immersed in a Static Staggered Field (SSF)
model turned out to be successful in describing the behaviour of
Haldane gap excitations as well.\cite{MZ98-L,Yokoo98} In agreement
with analytical\cite{MZ98-L,Bose99} and numerical\cite{numerics}
predictions, the energy gap was shown to increase quadratically
with $H_{\pi}$ in \pr, \nd, and \ndy\ systems in their
magnetically ordered phases.

The focus of the present paper is on a different, potentially very important,
aspect in the magnetism of \rr\ materials, namely on the strong spin-orbit
interactions typically associated with rare earth ions. Of particular interest
are local (single-ion) crystal field (CF) excitations  of the rare earths and
their possible interactions with the Haldane-gap modes propagating on the
Ni-chains. Until now this effect has not been investigated in sufficient
detail. The models used for $R^{3+}$ in \rr\ were in most cases restricted to
considering the ion's lowest-energy electronic configurations: effective
$S=1/2$ doublets for Kramers ions like Nd$^{3+}$ or
Er$^{3+}$,\cite{ZM98NBANO-L}  or singlets for Pr$^{3+}$
(Ref.~\onlinecite{Zheludev96PBANO}). Higher-energy electronic states were
entirely ignored. Yet, in a number of materials, particularly in
\nd,\cite{Zheludev96NBANO} several CF transitions occur very close to the
Haldane gap energy. In this situation a strong coupling between the two types
of excitations can be expected. Does this interaction limit the applicability
of the SSF model? Does it give rise to peculiar ``mixed'' Ni-$R$ excitations?
If so, do these ``mixed'' modes retain a predominantly 1-D character of Haldane
gap excitations? In the present paper we address these important and
interesting questions in a new series of inelastic neutron scattering
measurements on \nd.

\section{Experimental}
The crystal structure of \rr\  compounds is discussed in detail
elsewhere.\cite{Garcia93s} \nd\ crystallizes in an orthorhombic unit cell,
space group {\it Immm}, $a \approx 3.8$~\AA,~$b\approx 5.8$~\AA,~and $c\approx
11.7$~\AA.\cite{Yokoo97NDY} The $S=1$ Ni chains are formed by orthorhombically
distorted NiO$_{6}$ octahedra that are lined up in the $(100)$ crystallographic
direction, sharing their apical oxygen atoms. The Nd sites are positioned in
between these chains, linking nearest-neighbor chains along the
crystallographic $b$-axis (Fig.~\ref{struct}). We shall denote the
corresponding exchange constant as $J_1$. In addition, one can expect
substantial coupling of each Nd$^{3+}$ ion to two Ni$^{2+}$ sites in a third
nearby chain (exchange constant $J_2$). The site symmetry for the rare earth
ions is sufficiently low, and the electronic ground state for Nd$^{3+}$ is a
Kramers doublet. The N\'{e}el temperature for \nd\ is $T_{\rm N}=48$~K. The
magnetic structure is as described in Ref.~\onlinecite{Garcia,Sachan94}.

The simplest technique used to locat  CF excitations is inelastic
neutron scattering on  powder samples. Such measurements were
previously performed on \nd\ for energy transfers up to 30~meV
using a 3-axis spectrometer.\cite{Zheludev96NBANO} To learn about
higher-energy CF transitions we carried out preliminary
experiments on the time-of-flight (TOF) spectrometer LAM-D,
installed at the spallation neutron source KENS in the High Energy
Accelerator Research (KEK). LAM-D implements the so-called
inversion geometry and is equipped with four fixed PG crystal
analyzers. The measured spectra are automatically $Q$-integrated
in a certain momentum range, that roughly corresponds to
1--2~\AA$^{-1}$ at $\hbar\omega=0$. Energy transfers from -2~meV
to 200~meV are accessible. Energy resolution at the elastic
scattering position is about 300~$\mu$eV and the relative
resolution $\delta(\hbar \omega)/\hbar \omega$ is fairly constant,
around 4\%, between $\hbar \omega=2$ and $60$~meV. In the
experiment we used roughly 10~g of \nd\ powder prepared by the
solid state reaction method. The sample environment was a standard
closed-cycle refrigerator.

Most of the data presented below were obtained on single-crystal
samples. These experiments were carried out at the High Flux
Isotope Reactor (Oak Ridge National Laboratory) on the HB-1
triple-axis spectrometer. We used the same assembly of three
co-aligned single crystals as in previous studies.\cite{Yokoo98}
The sample was in all cases mounted with the $(h,k,0)$
reciprocal-space plane parallel to the scattering plane of the
spectrometer, to compliment previous $(h,0,l)$ measurements. A
neutron beam of $14.7$~meV or $30.5$~meV fixed final of $30.5$~meV
fixed-incident energy was used with $60'-40'-40'-120'$
collimations and a pyrolitic graphite (PG) filter positioned after
the sample. PG (002) reflections were used for monochromator and
analyzer. The measurements were done with energy transfers up to
35~meV. The sample environment was a standard Displex
refrigerator, and the temperature range 35--65~K was explored. The
growth directions  of the crystals roughly coincide with the
$(100)$ crystallographic axis. In the experiment the crystal rods
were therefore positioned horizontally in the scattering plane. In
this geometry absorption effects, primarily due to the presence of
Nd nuclei, may be substantial, as will be discussed in detail
below.

\section{Results}
\subsection{Powder data.}

Inelastic powder experiments, while obviously less informative than
single-crystal measurements, provide a convenient view of the excitation
spectrum in a wide energy range. Figures~\ref{powder}a-d show the measured
inelastic spectra for \nd\ powder at $T=16.5$, 55, 100, 300~K, respectively.
The background originating from the sample holder was measured separately and
subtracted from the data sets shown. Each scan corresponds to 16--20 hours of
counting time. At $T=55$~K, above the N\'{e}el temperature $T_{\rm N}=48~K$,
one clearly sees at least 3 features that are likely to be Nd-CF excitations
(Fig.~\ref{powder}b). In addition to the two previously identified low-energy
modes at $E_1=18$~meV and $E_2=24$~meV energy transfer,\cite{Zheludev96NBANO} a
broad yet intense peak is  seen at 40~meV. As in previous 3-axis powder
measurements, at 11~meV one can also see a shoulder corresponding to  Haldane
excitations. The latter are known to have a steep dispersion along the
chain-axis. As a result, compared to the dispersionless CF excitations, the
Ni-chain modes appear substantially weakened in a powder experiment, as opposed
to single crystal measurements. No substantial scattering is seen within the
Haldane gap. At $T=100$~K the 11~meV shoulder is much less pronounced, while
the CF peaks remain almost unchanged (Fig.~\ref{powder}c). As could be
expected, at high temperatures the entire spectrum becomes smeared out
(Fig.~\ref{powder}d).

A drastic change in the spectrum occurs upon cooling through the N\'{e}el
temperature (Fig.~\ref{powder}a, $T=16.5$~K). An intense resolution-limited
peak emerges inside the gap and is centered at 4~meV energy transfer. This
feature corresponds to a dispersion-less Ising-like spin wave associated with
long-range magnetic ordering.\cite{Yokoo98} Below $T=20$~K the Haldane gap was
previously shown to exceed 15~meV.\cite{Yokoo98} In apparent contradiction with
this single-crystal result, the 11~meV shoulder is weakened but still visible
in our new $16.5$~K powder data. One possible explanation of this is that the
shoulder is partially due to a weak, previously unobserved CF excitation: at
low temperatures the Haldane gap increases, but the new CF mode stays at 11~meV
energy transfer. Some of the single-crystal results presented below support
this interpretation.

A more complete, quantitative, analysis of the powder data is not presently
possible, as the phonon background is not known. In the future we plan to
supplement these data with ``background'' measurements on Y$_2$BaNiO$_5$, to
isolate the magnetic contribution of the rare earth ions in \nd. For the
purpose of the present paper however, we can draw one useful
 conclusion: in
trying to understand the coupling of the Ni-and Nd-subsystems on energy scales
comparable to $\Delta$, we can safely restrict ourselves to considering the two
strong CF excitations at 18~meV and 24~meV only.

\subsection{Transverse intensity modulation.}
\subsubsection{Constant-$Q$ scans}
Having verified the location of major CF excitations in powder measurements, we
proceeded with more detailed single-crystal experiments. As a first step, we
investigated the $b$-axis dependence of intensities and energies for the
Ni-chain- and Nd-CF-excitations near the 1-D AF zone-center. Typical
constant-$Q$ scans collected at $\bbox{Q}=(0,k,2.5)$, for several values of
$k$, are shown in Fig.~\ref{exdata}. A total of 9 scans of this type were
measured, with $k$ ranging from 0 to 1, at $T=55$~K (just above the temperature
of magnetic ordering, $T_{N}=48$~K). These data are summarized in the
intensity-contour plot in Fig.~\ref{tranmap1}a. Peaks corresponding to the
Ni-chain excitations and to the two CF modes are clearly seen. The central
result of this work is that {\it the relative intensities of these excitations
appear to vary significantly with transverse momentum transfer}.

To better understand this intensity variation, and to look for a transverse
dispersion in any of the three modes, we analyzed each of the measured
const-$Q$ scans using  a semi-empirical parameterized model cross section:
\begin{eqnarray}
 S(\bbox{Q},\omega) & = &
 S_H(\bbox{Q},\omega)+S_1(\bbox{Q},\omega)+S_2(\bbox{Q},\omega),\label{s1}\\
 S_H(\bbox{Q},\omega) & = & S^{(0)}_H\frac{\Delta^2}{\Gamma}\frac{1}
 {(\hbar\omega)^2+\Delta^2(\hbar\omega-\hbar\omega_{\bbox{Q}})^2/\Gamma^2}(f_{\rm{Ni}}(Q))^2,\label{s2}\\
 S_1(\bbox{Q},\omega) & = & S^{(0)}_1
 \delta(\hbar\omega-E_1)(f_{\rm{Nd}}(Q))^2,\label{s3}\\
 S_2(\bbox{Q},\omega) & = & S^{(0)}_2
 \delta(\hbar\omega-E_2)(f_{\rm{Nd}}(Q))^2,\label{s4}\\
 (\hbar\omega_{\bbox{Q}})^2 & = & \Delta^2 + v^2\sin^2(Q_{\|}a).\label{s5}
\end{eqnarray}
In these equations $S_H(\bbox{Q},\omega)$, $S_1(\bbox{Q},\omega)$,
and $S_2(\bbox{Q},\omega)$ are the partial contributions of the
Haldane-gap and two CF modes, respectively.  The coefficients
$S^{(0)}_H$, $S^{(0)}_1$, and $S^{(0)}_2$ represent the
intensities of each component. The spin wave velocity $v=
6.07\Delta$ \onlinecite{Sorensen94} and intrinsic energy width
$\Gamma$ characterize the Haldane-gap excitations. In our analysis
$\Gamma$ was fixed to a value of $1$~meV, as in  previous
studies.\cite{Yokoo98} $Q_{\|}$ is the projection of the
scattering vector onto the chain axis. Finally, $f_{\rm{Ni}}(Q)$
and $f_{\rm{Nd}}(Q)$ are magnetic form factors for Ni$^{2+}$ and
Nd$^{3+}$ ions.  The data analysis was performed assuming 7
independent parameters: the three intensity prefactors, a constant
background, and the three characteristic energies  $\Delta$, $E_1$
and $E_2$. The cross section was convoluted with the spectrometer
resolution function and fit separately to each scan. Typical fits
are shown in solid lines in Fig.~\ref{exdata}. This analysis
yielded a measurement of the $b$-axis dispersion relation in each
branch, plotted with symbols in Fig.~\ref{tranmap1}a. To within
experimental error the CF branches appear flat, and only a very
small dispersion is seen in the Haldane modes
(Fig.~\ref{results1}b). The Haldane gap energy is a maximum at the
transverse zone-boundary $k=0.5$. The intensity variation for each
mode is shown in Fig.~\ref{results1}b. The intensity of the 18~meV
excitation appears almost $k$-independent. At the same time,
intensities of both the Haldane- and the 24~meV CF-modes are
strongly $k$-dependent, and are seemingly in ``antiphase'' one to
another.

Data sets similar to those described above were also collected at $T=35$~K,
i.e., below the N\'{e}el temperature. These results are summarized in
Figs.~\ref{tranmap2} and \ref{results2}. Compared to the high-temperature
measurements, the Haldane gap energy is increased to about 14~meV. The Haldane
branch thus occurs closer to the 18~meV CF mode. It is also almost twice as
weak as at $T=55$~K. As a result, the scattering of data points in the measured
transverse dispersion curve shown in Fig.~\ref{results2}b is rather large, and
the bandwidth can not be reliably determined. Compared to the high-temperature
regime, the magnitude of intensity variations in the both the Haldane- and
24~meV- excitations are visibly reduced (see Fig.~\ref{results2}a).

\subsubsection{Global fits to data}
The final analysis of the measured constant-$Q$ scans was
performed by fitting a model cross section simultaneously to all
data collected at each temperature (Figs.~\ref{tranmap1}a and
\ref{tranmap2}a). The formulas \ref{s1}--\ref{s5} used to analyze
individual scans were modified to allow for a transverse
dispersion and intensity modulation in each branch. The
appropriate forms for these $k$-dependencies are derived in the
theory section below. The corresponding expression for the
magnetic dynamic structure factor near the 1-D AF zone-center is
as follows:

\begin{eqnarray}
 S(\bbox{Q},\omega) & = &
 S_H(\bbox{Q},\omega)+S_1(\bbox{Q},\omega)+S_1(\bbox{Q},\omega),\label{ss1}\\
 S_H(\bbox{Q},\omega) & = & S^{(0)}_H\frac{\Delta^2}{\Gamma}\frac{1}
 {(\hbar\omega)^2+\Delta^2(\hbar\omega-\hbar\omega_{\bbox{Q}})^2/\Gamma^2)}\left[(f_{\rm{Ni}}(Q))^2+A_H(f_{\rm{Ni}}(Q)f_{\rm{Nd}}(Q))\cos(Q_{\bot}b/2)\right],\label{ss2}\\
 S_1(\bbox{Q},\omega) & = & S^{(0)}_1
 \delta\left(\hbar\omega-E_1(Q_{\bot})\right)\left[(f_{\rm{Nd}}(Q))^2+A_1(f_{\rm{Ni}}(Q)f_{\rm{Nd}}(Q))\cos(Q_{\bot}b/2)\right],\label{ss3}\\
 S_2(\bbox{Q},\omega) & = & S^{(0)}_2
 \delta\left(\hbar\omega-E_2(Q_{\bot})\right)\left[(f_{\rm{Nd}}(Q))^2+A_2(f_{\rm{Ni}}(Q)f_{\rm{Nd}}(Q))\cos(Q_{\bot}b/2)\right],\label{ss4}\\
 (\hbar\omega_{\bbox{Q}}) & = & \sqrt{\Delta^2 +
 v^2\sin^2(Q_{\|}a)}+ B_H\cos^2(Q_{\bot}b/2),\label{ss5}\\
 E_1(Q_{\bot}) & = & E_1 + B_1\cos^2(Q_{\bot}b/2),\label{ss6}\\
 E_2(Q_{\bot}) & = & E_2 + B_2\cos^2(Q_{\bot}b/2).\label{ss7}
\end{eqnarray}
In these equations $Q_{\bot}$ is the component of momentum
transfer parallel to the $b$ axis. As explained in the theory
section, the dimensionless coefficients $A_H$, $A_1$ and $A_2$
quantify the admixture of Nd-CF spin fluctuations to the chain
modes and vice versa.  $B_H$, $B_1$ and $B_2$ are respective
magnitudes of transverse dispersion. To minimize the number of
adjustable parameters the values $B_1\equiv 0$, $A_2\equiv 0$ (no
dispersion observed) ,and $A_1\equiv 0$ (the mode appears not to
be modulated) were fixed, which hardly affected the final
$\chi^2$. The results of the global least-squares fitting to the
$T=55$~K and $T=35$~K data are shown in Figs.~\ref{tranmap1}b and
\ref{tranmap2}b, respectively. The refined values of all
parameters are summarized in Table~\ref{results}. In both cases a
$\chi^2=1.7$ was achieved, which suggests that the model cross
section describes the observed behavior reasonably well. As
discussed above, the coefficients for intensity modulations in all
three modes, as well as the dispersion bandwidth for the Haldane
branch at $T=55$~K, determined in this analysis, can be considered
rather reliable. On the contrary, the large value for $B_{\rm H}$
at $T=35$~K is somewhat suspect. In the refinement this parameter
has a strong correlation with $E_1$, i.e., the Haldane gap is not
very well resolved. This also accounts for the large scattering in
the points obtained in fits to individual scans in
Fig.~\ref{results2}c. There is little doubt however that the
bandwidth of transverse dispersion remains very small even in the
magnetically ordered state.

\subsubsection{Wide-range constant-$E$ scans}
It is implied in Eqs.~\ref{ss2}--\ref{ss4} that the intensity modulation in all
modes is periodic along the $b$-axis. To verify this periodicity we performed
constant-$E$ scans at the Haldane gap energy and at $\hbar \omega=25$~meV,
covering a broad range of transverse momentum transfers. Typical raw data
collected at $T=55$~K are shown in Figs.~\ref{exabs1}a and \ref{exabs2}a. A
serious technical problem with such measurements is that they are strongly
influenced by neutron absorbtion in the sample. In each scan $k$ varies
significantly at a constant value of $h$. As a result, the crystal rods rotate
relative to the incident and outdoing beams by a large angle. Absorbtion
corrections could be calculated, but additional undesirable effects arising
from the size (length) of the sample being comparable to the width of the
incident neutron beam, are difficult to account for analytically. To correct
the data for any transmission effects, we measured the actual transmission
coefficient {\it in situ}. This was done separately for each point of the
constant-$E$ scans performed. The monochromator, sample and scattering angles
of the spectrometer were set to values calculated for a given point in the
inelastic scan. The analyzer angles were then set to the elastic position to
measure the intensity of incoherent scattering. Incoherent scattering being
isotropic, the measured incoherent intensity can be assumed to be proportional
to the effective transmission coefficient. Typical measured transmission curves
are shown in Figs.~\ref{exabs1}b and \ref{exabs2}b. A few Bragg reflections
were accidentally picked up in these elastic scans, and the corresponding
points were removed from the data sets shown. In Figs.~\ref{exabs1}b and
\ref{exabs2}b arrows indicate geometries of minimum transmission, realized when
the sample rods are parallel to the incident or the scattered beams. The
measured transmission coefficients were used to scale the measured const-$E$
scans point-by-point. The resulting corrected data are plotted in
Figs.~\ref{haldanemodulation} and \ref{cfmodulation}.

Figure~\ref{haldanemodulation} shows transmission-corrected $k$-scans taken at
the Haldane gap energy, near two 1-D AF zone-centers $h=1.5$ and $h=2.5$. The
intensity fall-off at large $|k|$ is mostly due to focusing and to the effect
of magnetic form-factors. On top of this fall-off, a periodic intensity
modulation is clearly seen. The data can be rather well fit to Eq.~\ref{ss2}
convoluted with the spectrometer resolution function. The fits are shown in
solid lines in Fig.~\ref{haldanemodulation} and correspond to $A_{\rm
H}=-0.19(0.02)$. This value is in good agreement with $A_{\rm H}=-0.17$,
obtained from the analysis of constant-$Q$ scans described above.

Figure~\ref{cfmodulation}b shows that near the 1-D AF zone-center
the intensity of the 25~meV CF mode is modulated as well, in
antiphase with that of the Haldane branch. The data can be fit to
Eq.~\ref{ss4}, assuming $A_2= 0.28(0.05)$ (solid line in
Fig.~\ref{cfmodulation}b). Note that this value is less than a
half of that deduced from the analysis of constant-$Q$ scans at
the same temperature ($A_2= 0.72$). This discrepancy is easily
understood. Even at 25~meV energy transfer the Haldane-gap
excitations contribute to inelastic scattering, thanks to their
steep dispersion along the chain axis that produces long ``tails''
on the high-energy side of the peak in constant-$Q$ scans. The
contribution of these tails was not taken into account in the
analysis of the wide-range const-$E$ scan shown in
Figure~\ref{cfmodulation}b, but was included in the global fit to
constant-$Q$ scans. Our previous estimate for $A_2$ should thus be
considered more accurate.

An important result is that away from the 1-D AF zone center, at $h=2.25$, the
25~meV mode shows no oscillations of intensity (Fig.~\ref{cfmodulation}a). To
within experimental error the measured $k$ dependence can be explained by the
Nd-form factor alone (solid line). This observation supports the notion that
intensity oscillations result from a mixing between CF and Haldane-gap
excitations. Away from the 1-D AF zone-center Ni-chain modes occur at very high
energy transfers and therefore do not interact with the dispersionless CF
branch.

\subsubsection{Previous data analysis procedures revised}
The discovery of a substantial variation of the intensity in the CF excitations
raises concerns about the procedure used to analyze the inelastic data for \nd\
in Refs.~\onlinecite{Yokoo98,Raymond99}. In these studies the CF ``background''
was measured away from the 1-D AF zone-centers, where the Haldane modes are not
seen. The background was then subtracted point-by-point from the ``signal''
scans measured at the 1-D AF zone-center. It was assumed that such
background-subtracted scans are a good measure of inelastic scattering
originating from the Ni-chains alone. We now understand that this approach is,
in principle, not valid. Although CF excitations are almost dispersionless,
their intensity in the ``background'' scans will not necessarily match that in
the ``signal'' scans. According to the theoretical model described below, the
procedure can only be applied at $k=(2n+1)/2$, $h=(2m+1)/2$ ($m$, $n$-
integer), where there is no mixing between Ni-chain and Nd-single-ion
excitations. The data analysis employed in Ref.~\onlinecite{Raymond99}, where
measurements were performed at $\bbox{Q}=(1.5,0.5,0)$, is thus fully
appropriate. In Ref.~\onlinecite{Yokoo98}, on the other hand, the data were
taken at $k=0$, and the background subtraction may, in principle, have produced
systematic errors. It appears however that point-by-point background
subtraction is a fairly reliable way to determine the Haldane gap {\it energy}
at any transverse wave vector. Indeed, in Ref.~\onlinecite{Yokoo98} all scans
were taken below 25~meV energy transfer, and thus include only the 18~meV CF
excitation. This mode, as we now know, is in fact flat. The main results of
Ref.~\onlinecite{Yokoo98} pertain to the temperature dependence of the Haldane
gap, that occurs even below 18~meV in the studies temperature range, are
totally valid.

\subsection{Dispersion along the chain axis}
The mixing between the Haldane-gap and 24~meV CF modes should be
most prominent at points of reciprocal space where the two
excitations have similar energies. While the CF excitation is
practically dispersionless, the Haldane branch has a steep
dispersion along the chain axis, and can be expected to anticross
with the 24~meV branch rather close to the 1-D AF zone-center. To
investigate this phenomenon we measured the intensity of inelastic
neutron scattering from \nd\ on a grid of points in $h-E$ space at
$k=0.5$, $h=1.5-1.8$, at $T=55$~K. The result of this measurement
is summarized in the grayscale/contour plot in Fig.~\ref{longmap}.
The circles and squares show the positions of peaks seen when the
data are broken up into a collection of constant-$Q$ and
constant-$E$ scans, respectively. The most prominent feature of
the spectrum is an anticrossing at the point of intersection of
the flat 24~meV CF mode and the parabolic Haldane branch. The
solid lines are guides for the eye that emphasize this effect. The
anticrossing gap $2\delta E$ can be estimated to be roughly 4~meV.
Note that there is no evidence of anticrossing of the Haldane and
18~meV-CF branches.

A much less significant feature that is barely seen within the statistics of
the measurement is what appears to be a new dispersionless mode at $\hbar
\omega \approx 12$~meV (dash-dot line in Fig.~\ref{longmap}). In the single
crystal sample it appears much weaker than the the Haldane mode, but
nevertheless may account for the ``shoulder'' seen in powder samples at
$T=16.75$~K, as discussed previously. This new feature is substantially weaker
than the other two CF excitations, and we shall disregard it in the following
analysis.

\section{Theory}
It is rather interesting that in \nd\ the pronounced transverse modulation of
intensity in the Ni-chain and CF excitations is {\it not} accompanied by any
significant transverse dispersion. If we were dealing with a one-component
system, e.g., {\it directly coupled} chains, a narrow bandwidth along a given
direction would automatically result in only a weak periodicity in the
structure factor. For magnetic excitations in a Bravais crystal, this follows
from the sum rule for the first moment of the magnetic dynamic structure factor
$S(q,\omega)$.\cite{Muller} The strong modulation and lack of dispersion in
\nd\ must therefore be an interference effect, that involves both Ni- and Nd
magnetic degrees of freedom. In the following sections we shall demonstrate
that this mixing can be qualitatively accounted for by a simple Random Phase
Approximation (RPA) model for Ni-$R$ interactions.

\subsection{Bare susceptibilities}
\subsubsection{Ni-chains}
The first step in the RPA analysis is to determine the bare (non-interacting)
dynamic susceptibilities for the Ni and Nd sublattices. For the  Haldane spin
chains in the vicinity of the 1-D AF zone-center $q_{\|}=\pi$ we can safely use
the single mode approximation \cite{Arovas88,Muller81,Ma92}:
\begin{equation}
 \chi^{(0)}_{\rm Ni}(\bbox{Q},\omega)=\frac{1-\cos Q_{\|}a}{2}\frac{Zv}{\Delta^2+v^2\sin^2(Q_{\|}a)-(\hbar \omega+i\epsilon)^2}
 \label{bareni}
\end{equation}
In this formula $Q_{\|}$ is the projection of the momentum
transfer $\bbox{Q}$ onto the chain axis, $v$ is the spin wave
velocity in the chains, $v\approx 6.07\Delta$, and $Z\approx 1.26$
(Ref.~\onlinecite{Sorensen94}). It is easily verified that for
scattering vectors close to the 1-D AF zone-center Eq. \ref{s2}
used to fit the the inelastic scans coincides with the imaginary
part of Eq.~\ref{bareni} in the limit $\epsilon\rightarrow 0$. In
the magnetically ordered state, {\it i.~e.}, in the presence of an
effective mean field $\bbox{H}_{eff}$, Eq.~(\ref{bareni}) can
still be utilized. However, the corresponding gap energies for the
three components of the Haldane triplet will be modified by
$\bbox{H}_{eff}$, as discussed in
Refs.~\onlinecite{ZM98NBANO-L,MZ98-L,MZ97,Yokoo98}.

\subsubsection{Rare earth ions}
For the rare earth subsystem the bare single-ion susceptibility can not be
written down so easily, as the exact electronic configuration of $R^{3+}$ in
\rr\ has not been calculated to date. When we were concerned with the mechanism
of magnetic ordering in \nd\  in Refs.\onlinecite{ZM98NBANO-L,Yokoo98}, we were
dealing with the {\it static} susceptibility of the rare earths. For Nd$^{3+}$
at low temperatures the static susceptibility can be, to a good approximation,
attributed to the lowest-energy Kramers doublet and written as a $S=1/2$
Brillouin function. For our present purpose of studying the {\it dynamic}
coupling of Nd and Ni subsystems we are interested in the {\it dynamic}
susceptibility of Nd$^{3+}$ at frequencies comparable to the Haldane gap
energy. In the paramagnetic phase the contribution to $\chi_{\rm Nd}$ from the
ground state doublet is {\it purely elastic} and can  thus be totally ignored
in the RPA calculation. At $T>T_{\rm N}$ we therefore only need to consider the
contribution of transitions between the ground state of the ion and
higher-energy crystal-field levels. In the ordered phase,  we also have to take
into account single-ion transitions within the ground state double, which
becomes split by the mean exchange field\cite{Yokoo98}. As a first step, we
shall consider only one excited level of Nd$^{3+}$, at energy $E$. Transitions
from the ground state to the excited state produce the following term in Nd
single-ion susceptibility:
\begin{equation}
\chi^{(0)}_{\rm Nd}(\omega)  =  M^2\frac {E} {E^2-(\hbar \omega +i\epsilon)^2
}R(T) \label{barend}
\end{equation}
In this formula $M$ is the matrix element of the magnetic moment
operator between the ground state and the excited state. $R(T)$ a
the temperature renormalization factor defined as the difference
in populations of the excited and ground states. To simplify our
qualitative analysis, in Eq.~\ref{barend} we have ignored any
terms that are off-diagonal in spin projection indexes, and the
above formula thus applies to a particular selected channel of
spin polarization.

\subsection{Interaction geometry}
Let us now consider the geometry of Ni-Nd exchange interactions
(Fig.~\ref{struct}). The antiferromagnetic coupling denoted by
$J_1$ stabilizes the magnetic structure seen in the
low-temperature phase, and is responsible for the staggered mean
field at $T<T_{\rm N}$. As is often the case in body-centered
crystals, the coupling $J_2$ is geometrically frustrated. Indeed,
it links a single $R^{3+}$ moment to two subsequent Ni$^{2+}$
spins in the chains, that are strongly coupled
antiferromagnetically between themselves. As a result, at the MF
level $J_2$ plays no role in long-range ordering. Below we shall
arbitrarily assume that the exchange matrix is diagonal in spin
projection indexes. In the body-centered structure, with 2
$R$-sites per every Ni, we have to consider 6 magnetic ions: Ni(1)
at (0.5,0.5,0), Ni(2) at (0,0,0.5), Nd(3) at (0.5,0,d), Nd(4) at
(0.5,0,-d), Nd(5) at (0,0.5,d-0.5), and Nd(6) at (0,0.5,-[d-0.5]),
where $d\approx 0.2$. For the RPA calculation we need the Fourier
transform of the exchange matrix in this basis set:
\begin{eqnarray}
J_{\bot}(\bbox{Q})=\left(
\begin{array}{cccccc}
 0 & 0 & J_1(\bbox{Q}) & J_1^\ast(\bbox{Q}) &
J_2(\bbox{Q}) & J_2^\ast(\bbox{Q})\\
 0 & 0 & J_2(\bbox{Q}) & J_2^\ast(\bbox{Q}) & J_1(\bbox{Q}) & J_1^\ast(\bbox{Q})\\
 J_1^\ast(\bbox{Q}) & J_2^\ast(\bbox{Q}) & 0
 & 0 & 0 & 0\\
 J_1(\bbox{Q}) & J_2(\bbox{Q}) & 0
 & 0 & 0 & 0\\
 J_2^\ast(\bbox{Q}) & J_1^\ast(\bbox{Q}) & 0
 & 0 & 0 & 0\\
 J_2(\bbox{Q}) & J_1(\bbox{Q})& 0
 & 0 & 0 & 0
 \end{array}\right)\\
 J_1(\bbox{Q})=2J_1\cos(\pi k)\exp(2 \pi i l d)\label{j1q}\\
 J_2(\bbox{Q})=2J_2\cos(\pi h)\exp[2 \pi i l (d-0.5)]
 \end{eqnarray}
One readily sees that, in the reciprocal-space planes $(h,\frac{2n+1}{2},l)$
($n$-integer), $J_1$-interactions cancel out and can be ignored. A similar
cancellation occurs for $J_2$ in the $(\frac{2n+1}{2},k,l)$ ($n$-integer)
planes. These cancellations result from interaction topology, and do not rely
on any of our assumptions regarding the bare susceptibilities.

\subsection{RPA susceptibility}
The assumptions made above bring us to a simple mode-coupling problem in each
channel of spin polarization. The RPA susceptibility matrix is written as
$\chi^{\rm RPA}(\bbox{Q},\omega)=\left[ 1+ \chi^{0}(\bbox{Q},\omega)J(\bbox{Q})
\right] ^{-1} \chi^{0}(\bbox{Q},\omega)$.

\subsubsection{Dispersion}
The dispersion relation is obtained by solving the equation $\det \left(\left[
\chi^{RPA} \right]^{-1}\right)=0$. In our case  of only one excited Nd-state
this can easily be done analytically. With 6 independent atoms we obtain 6
modes (two of them degenerate):
\begin{eqnarray}
 2 (\hbar \omega_{1,2})^2 & = &
 \Delta_{Q_{\|}}^2+E^2\pm\sqrt{(\Delta_{Q_{\|}}^2-E^2)^2+8EM^2Zv{\cal J}_+^2} \nonumber \\
 2 (\hbar \omega_{3,4})^2 & = &
 \Delta_{Q_{\|}}^2+E^2\pm\sqrt{(\Delta_{Q_{\|}}^2-E^2)^2+8EM^2Zv{\cal J}_-^2} \nonumber \\
 \hbar \omega_{5,6} & = & E,\label{goode}
\end{eqnarray}
where we have introduced the notations
$\Delta_{Q_{\|}}^2=\Delta^2+v^2\sin^2(Q_{\|}a)$, and
\begin{equation}
{\cal J}_{\pm}^2=\left|J_1(\bbox{Q})\pm J_2(\bbox{Q})\right|^2=4J_1^2\cos^2(\pi
k)+4J_2^2\cos^2(\pi h)\pm 16J_1J_2\cos(\pi k)\cos(\pi h)\cos(\pi l).
\end{equation}
We have implicitly dropped the $(1-\cos Q_{\|}a)/2$ factor in $\chi^{(0)}_{\rm
Ni}$ and the temperature factor $R$ in  $\chi^{(0)}_{\rm Nd}$.

Particularly useful is the linearized version of Eq.~\ref{goode} which becomes
valid in the weak coupling limit:
\begin{eqnarray}
 \hbar \omega_1({\bf Q})& = & \Delta_{ Q_{\|} }-{\cal J}_+(\bbox{Q})^2\frac{Zv}{\Delta}\frac{EM^2 }{E^{2}-\Delta_{ Q_{\|} }^2},\nonumber \\
 \hbar \omega_2({\bf Q})& = & E+{\cal J}_+(\bbox{Q})^2\frac{Zv}{\Delta}\frac{EM^2 }{E^{2}-\Delta_{ Q_{\|} }^2},\nonumber \\
 \hbar \omega_3({\bf Q})& = & \Delta_{ Q_{\|} }-{\cal J}_-(\bbox{Q})^2\frac{Zv}{\Delta}\frac{EM_1^2 }{E^{2}-\Delta_{ Q_{\|} }^2},\nonumber \\
 \hbar \omega_4({\bf Q})& = & E+{\cal J}_-(\bbox{Q})^2\frac{Zv}{\Delta}\frac{EM_1^2 }{E^{2}-\Delta_{ Q_{\|} }^2},\nonumber \\
 \hbar \omega_{5,6}({\bf Q})& = & E,\label{lindisp}
\end{eqnarray}
The 1st and 3rd branches can be described as Ni-chain Haldane gap modes with an
admixture of Nd-spin correlations. The 2nd and 4th branches, conversely, are CF
excitations with a small component of Ni-spins. If the Nd-single-ion excitation
lies above the Haldane branch at a given wave vector, the energy of the
Ni-chains mode is pushed down by Ni-Nd interactions, while the CF excitation
energy is increased. The reverse is expected in the case of a single-ion
excitation inside the Haldane gap.  Modes 5 and 6 appear as totally unperturbed
single-ion transitions. A very important result is that the transverse
dispersion resulting from Ni-$R$ interactions is {\it quadratic} in ${\cal J}$
and is thus expected to be very small in the weak-coupling limit.

Also useful for our purposes are values of anticrossing gaps at the point of
mode intersection. According to Eqs.~\ref{goode} these are given by:
\begin{eqnarray}
 2 \delta E_{1\leftrightarrow2} & = &
 {\cal J}_+M\sqrt{\frac{2Zv}{E}} \nonumber\\
 2 \delta E_{3\leftrightarrow4} & = &
 {\cal J}_-M\sqrt{\frac{2Zv}{E}},\label{landaugap}
\end{eqnarray}
for the two pairs of intersecting same-symmetry branches, respectively.
Intersection of branches of different symmetry (1,4) and (2,3) does not lead to
an anticrossing effect.

\subsubsection{Intensity modulation}
The inelastic neutron scattering cross section is related to the imaginary part
of dynamic susceptibility through the fluctuation-dissipation theorem. To get
the actual intensities we also have to take into account the Ni- and $R$-
magnetic form factors $f_{\rm Ni}(Q)$ and $f_{\rm Nd}(Q)$:
\begin{equation}
\frac{d \sigma}{d \Omega}\propto \left(\begin{array}{cccccc}
 f_{\rm Ni}(Q) & f_{\rm Ni}(Q)& f_{\rm Nd}(Q) & f_{\rm Nd}(Q) & f_{\rm Nd}(Q) & f_{\rm Nd}(Q)
\end{array}\right) \chi''(\bbox{Q},\omega) \left(\begin{array}{c}
f_{\rm Ni}(Q)\\ f_{\rm Ni}(Q)\\ f_{\rm Nd}(Q)\\ f_{\rm Nd}(Q)\\ f_{\rm Nd}(Q)\\
f_{\rm Nd}(Q)
\end{array}\right)
.\label{trick}
\end{equation}
The task of calculating the total $\chi''(\bbox{Q},\omega)$ is greatly
simplified in the weak-coupling limit, where one can write $\chi^{\rm
RPA}(\bbox{Q},\omega) \approx \left[ 1- \chi^{0}(\bbox{Q},\omega)J(\bbox{Q})
\right]\chi^{0}(\bbox{Q},\omega)$. At this level of approximation the two
Haldane excitations (modes 1 and 3) are not distinguished. The same applies to
CF modes 2,4,5,6. For the sum of partial cross section of the Haldane modes a
straightforward yet somewhat tedious calculation gives:
\begin{equation}
 \frac{d \sigma_{1+3}}{d \Omega} \propto \frac{Zv}{\Delta_{Q_{\|}}}\left(f_{\rm Ni}(Q)^2-4f_{\rm
 Ni}(Q)f_{\rm Nd}(Q)\frac{EM^2}{E^2-\Delta_{Q_{\|}}^2}\mbox{Re}\left[J_1(\bbox{Q})+J_2(\bbox{Q})\right]\right).\label{halmod}
\end{equation}
Similarly, for the single-ion modes:
\begin{equation}
\frac{d \sigma_{2+4}}{d \Omega} \propto {2M^2}\left(f_{\rm
Nd}(Q)^2+4f_{\rm
 Ni}(Q)f_{\rm Nd}(Q)\frac{Zv}{E^2-\Delta_{Q_{\|}}^2}\mbox{Re}\left[J_1(\bbox{Q})+J_2(\bbox{Q})\right]\right)\label{cfmod}.
\end{equation}
One readily sees that, unlike the corrections to the dispersion relation,
corrections to mode intensities are {\it linear} in $J_1$ and $J_2$ and can be
substantial. The linear form of Eqs.~\ref{lindisp},\ref{halmod} and \ref{cfmod}
enables a straightforward generalization to the case of several Nd-single-ion
transitions.

\subsubsection{Results for the 1-D AF zone-center}
The expressions for the dynamic structure factor can be further simplified for
scattering vectors in $(\frac{2n+1}{2},k,l)$ ($n$-integer) reciprocal-space
planes, where most of the neutron data were collected. Here the effect of $J_2$
is totally cancelled, and there are only 3 branches in the spectrum: the
Haldane branch (labeled ``H''), a CF branch that interacts with the Ni-chain
excitations (``CF''), and the unperturbed CF transition (``CF$_0$''). Using
Eq.~\ref{trick} on the exact (rather than linearized) RPA susceptibility
matrix, and truncating the result to the first order in $J_1$ we obtain:
\begin{eqnarray}
 \frac{d \sigma_{\rm H}}{d \Omega}& \propto & \frac{Zv}{\Delta_{Q_{\|}}}\left(f_{\rm Ni}(Q)^2-8J_1f_{\rm
 Ni}(Q)f_{\rm Nd}(Q)\frac{EM^2}{E^2-\Delta_{Q_{\|}}^2}\cos(\pi k)\cos(2\pi l
 d)\right),\label{halintresult} \\
\frac{d \sigma_{\rm CF}}{d \Omega} & \propto & {2M^2}\left(f_{\rm Nd}(Q
)^2[1+\cos(4\pi l d)]+8J_1f_{\rm
 Ni}(Q)f_{\rm Nd}(Q)\frac{Zv}{E^2-\Delta_{Q_{\|}}^2}\cos(\pi k)\cos(2\pi l
 d)\right),\label{m1intresult} \\
 \frac{d \sigma_{{\rm CF}_0}}{d \Omega} & \propto & {2M^2}f_{\rm Nd}(Q)^2[1-\cos(4\pi l
 d)].
\end{eqnarray}
The intensity of all 3 modes become modulated. The period of modulation along
the $b$-axis is exactly $2b^{\ast}$, while that along $c$ is incommensurate
with the lattice.  For the reciprocal-space rods $(\frac{2n+1}{2},k,0)$
($n$-integer) the mode CF$_{0}$ vanishes and only 2 branches remain. For the
purpose of convenience we shall also re-write Eqs.~\ref{lindisp} for the
$(\frac{2n+1}{2},k,l)$ ($n$-integer) planes:
\begin{eqnarray}
 \hbar \omega_{\rm H}({\bf Q})& = & \Delta_{ Q_{\|} }-4J_1(\bbox{Q})^2\frac{Zv}{\Delta}\frac{EM^2 }{E^{2}-\Delta_{ Q_{\|} }^2}\cos^2(\pi k), \label{haldispresult}\\
 \hbar \omega_{\rm CF}({\bf Q})& = & \Delta_1+4J_1(\bbox{Q})^2\frac{Zv}{\Delta}\frac{EM^2 }{E^{2}-\Delta_{ Q_{\|} }^2}\cos^2(\pi k), \\
 \hbar \omega_{{\rm CF}_0}({\bf Q})& = & E.\label{emod}
\end{eqnarray}
As noted previously, the magnitude of transverse dispersion is proportional to
$J_1^2$.

We can not expect the model described above to be quantitatively valid for any
real \rr\ system. Its main limitation is that it assumes the exchange matrix
and the matrix elements of momentum between the rare earth ground- and excited
states to be simultaneously diagonal in spin projection indexes. In the \rr\
structure, the rare earth sites have very low symmetry and the bare
susceptibility and exchange matrixes are bound to be rather complex. We can
however expect the model to make qualitatively correct predictions regarding
the periods and signs of modulations of excitation intensities and energies.

\subsubsection{Paramagnetic phase}
For \nd\ we should distinguish two regimes. At $T>T_{\rm N}$, at energy
transfers below 30~meV, only the 18~meV and the 24~meV Nd-modes will mix with
the Haldane-gap excitations. Experimentally, the 18~meV transition appears
totally decoupled from the Ni-chains . This is most likely due to its
particular polarization and the actual form of the exchange tensors. For
example, if the 18~meV excitation were polarized along the $x$ axis, and
$J_{1}^{xx}$ is zero, than no mixing would occur. The 24~meV excitation is thus
the only one that mixes with the Haldane branches, producing the observed
modulations.

The RPA model, oversimplified as it is, justifies the use of
Eqs.~\ref{ss1}--\ref{ss7} in fitting the inelastic neutron
scattering data. From this analysis we can even extract some
useful quantitative information. According to
Eqs.~\ref{haldispresult} and \ref{halintresult}, the ratio $B_{\rm
H}/A_{\rm H}$ (see Eqs.~\ref{ss2},\ref{ss5}) is {\it independent}
of any parameters of the rare earth ions:
\begin{equation}
B_{\rm H}/A_{\rm H}=\frac{Zv}{2\Delta}J_1
\end{equation}
This is a robust result that allows us to obtain at least an order-of-magnitude
estimate for $J_1$, without making any assumptions regarding Nd$^{3+}$. For an
isolated Haldane spin chain $Zv\approx 7.6\Delta$, so $B_{\rm H} /A_{\rm
H}\approx 3.8 J_{1}$. From the $T=55$~K experimental values $A=-0.17$~meV and
$B=-0.19$ we get $J_{1}\approx 0.3$~meV. We can compare this value to the MF
coupling $\alpha$ for \nd\ that we obtained in
Refs.~\onlinecite{ZM98NBANO-L,Yokoo98}. Indeed, $J_{\bot}$ should be related to
$\alpha$ through $J_{1}M^{(Ni)}M^{(Nd)}=\alpha/2$, which gives $J_{1}\approx
0.5$~meV, in reasonable agreement with our new estimate.

A more crude estimate can also be obtained  for $J_2$, which in our model is
the only possible source of the observed anticrossing between the Haldane and
24~meV CF branches (Fig.~\ref{longmap}). Indeed, at $k=0.5$ where these data
were taken the effect of $J_1$ is cancelled, according to Eq.~\ref{j1q}.
Experimentally, the anticrossing is about 4~meV. From Eq.~\ref{landaugap} we
get $J_2M_1\approx 0.8$~meV. The 24~meV CF mode and the Haldane excitation are
of comparable intensity, and based on Eqs.~\ref{halintresult} and
\ref{m1intresult} we can assume that $M^2\sim Zv/\Delta$, which gives $M\sim
3$. This gives us an estimate for $J_2$: of the order of 0.2~meV. This is a
rather reasonable value for a rare earth- transition metal superexchange bond.

\subsubsection{Ordered state}
In the magnetically ordered phase an additional Nd-centered excitation, namely
the 4~meV Ising-like spin wave that appears below $T_{\rm N}$
(Fig.~\ref{powder}a) will also mix with the Ni-chain modes. Being inside the
Haldane gap, it will produce intensity and energy modulations that are opposite
to those resulting from the mixing with the high-energy 24~meV branch.
Intensity oscillations in the Haldane branch will thus be reduced, in agreement
with our experimental findings. As observed in previous studies, the intensity
of the peak corresponding to Haldane excitations is reduced by almost a factor
of 2. This results from the suppression of the $b$-axis-polarized Haldane
excitation (longitudinal mode) in the ordered state.\cite{Raymond99} In the
framework of the RPA model, a decrease of spectral weight in the Haldane
branches inevitably leads to a smaller interference effect in the CF mode. Such
a reduction of intensity modulation in the ordered state is indeed observed
experimentally for the 24~meV excitation (see Table~\ref{results}).

\section{Summary}
We have demonstrated that local excitations associated with magnetic rare earth
ions play an important role in the magnetism of \rr\ quantum antiferromagnets.
Even in the 1-D (paramagnetic) phase there are substantial {\it dynamic}
interactions between quantum spin chains and the rare earth subsystem. This
phenomenon can be qualitatively accounted for by a simple chain-RPA model. Not
having more detailed information on the electronic states of the rare earth
ions, using this model for a quantitative analysis of the data is obviously a
leap of faith. It does however produce reasonable order-of-magnitude estimates
for Ni-Nd coupling constants.

At any temperature the Haldane-gap modes can not be considered as
purely-Ni excitations, and their interference with Nd-single-ion
modes is substantial. Despite that, the Haldane gap {\it energy}
is practically the same as in {\it isolated} chains in a {\it
static} mean exchange field, and the transverse dispersion is very
small. The Static Staggered Exchange Field  model is thus fully
justified for \nd, as are all our previous results for the
temperature- (staggered field-) dependence of the Haldane gap.

\acknowledgements The authors would like to thank P. Dai, J. Zarestky, B.
Taylor and R. Rothe for expert technical assistance. Work at BNL was carried
out under Contract No. DE-AC02-98CH10886, Division of Material Science, U.S.
Department of Energy. Oak Ridge National Laboratory is managed for the U.S.
D.O.E. by Lockheed Martin Energy Research Corporation under contract
DE-AC05-96OR22464.


\begin{thebibliography}{10}

\bibitem{Zheludev99NN}
For a recent review see A. Zheludev, Neutron News {\bf 10}, No. 3, p. 16
  (1999).

\bibitem{YBANO}
J. Darriet and L. P. Regnault, Solid State Commun., {\bf 86}, 409 (1993); T.
  Yokoo, T. Sakaguchi, K. Kakurai and J. Akimitsu, J. Phys. Soc. Japan {\bf
  64}, 3651, (1995); G. Xu {\it et al.}, Phys. Rev. B {\bf 54}, R6827 (1996).

\bibitem{Zheludev96PBANO}
A. Zheludev, J. M. Tranquada, T. Vogt, D. J. Buttrey, Europhys. Lett. {\bf 35},
  385 (1996); Prys. Rev. B {\bf 54}, 6437 (1996).

\bibitem{Zheludev96NBANO}
A. Zheludev, J.~M. Tranquada, T. Vogt, and D.~J. Buttrey, Phys. Rev. B {\bf
  54},  7216  (1996).

\bibitem{Yokoo97NDY}
T. Yokoo, A. Zheludev, M. Nakamura, and J. Akimitsu, Phys. Rev. B {\bf 55},
  11516  (1997).

\bibitem{Yokoo98}
T. Yokoo {\it et al.}, Phys. Rev. B {\bf 58}, 14424 (1998).

\bibitem{Raymond99}
S. Raymond {\it et~al.}, Phys. Rev. Lett. {\bf 82},  2382  (1999).

\bibitem{Garcia}
E. Garc\'{i}a-Matres {\it et al.}, J. Solid State Chem. {\bf 103}, 322 (1993);
  J. Mag. Magn. Mater. {\bf 149}, 363 (1995).

\bibitem{ZM98NBANO-L}
A. Zheludev {\it et~al.}, Phys. Rev. Lett. {\bf 80},  3630  (1998).

\bibitem{MZ98-L}
S. Maslov and A. Zheludev, Phys. Rev. Lett. {\bf 80},  5786  (1998).

\bibitem{numerics}
J. Lou, X. Dai, S. Quin, Z. Su and L. Yu, Phys. Rev. B {\bf 60},
52 (1999).

\bibitem{Bose99}
I. Bose and E. Chattopadhyay, cond-mat/9904311.

\bibitem{Garcia93s}
E. Garc\'{i}a-Matres {\it et~al.}, J. Solid State Chem. {\bf 103},  322
  (1993).

\bibitem{Sachan94}
V. Sachan, D.~J. Buttrey, J.~M. Tranquada, and G. Shirane, Phys. Rev. B {\bf
  49},  9658  (1994).

\bibitem{Sorensen94}
E. S. Sorensen and I Affleck, Phys. Rev. B{\bf 49}, 15771 (1994).

\bibitem{Muller}
G. Muller, H. Thomas, M. W. Puga, and H. Beck, J. Phys. C: Solid State Phys.
  {\bf 14}, 3399 (1981).

\bibitem{Arovas88}
D.~P. Arovas, A. Auerbach, and F.~D.~M. Haldane, Phys. Rev. Lett. {\bf 60},
  531  (1988).

\bibitem{Muller81}
G. Muller, H. Thomas, M.~W. Puga, and H. Beck, J. Phys. C {\bf 14},  3399
  (1981).

\bibitem{Ma92}
S. Ma {\it et~al.}, Phys. Rev. Lett. {\bf 69},  3571  (1992).


\bibitem{MZ97}
S. Maslov and A. Zheludev, Phys. Rev. B {\bf 57},  68  (1998).

\end{thebibliography}

\begin{figure}
\caption{A schematic representation of the magnetic interaction geometry in the
structure of \rr. } \label{struct}
\end{figure}

\begin{figure}
\caption{LAM-D inelastic spectra measured in \nd\ powder samples at different
temperatures. The contribution of the sample holder has been subtracted. }
\label{powder}
\end{figure}

\begin{figure}
\caption{Typical constant-$Q$ scans measured in \nd\ at $T=55$~K at the 1-D AF
zone-center $h=2.5$, at different momentum transfers perpendicular to the chain
axis (symbols). The solid lines are a global fit to the model cross section, as
described in the text. The hatched, light-gray and gray peaks represent partial
contributions from the Haldane-gap and two crystal-field excitations,
respectively.} \label{exdata}
\end{figure}

\begin{figure}
\caption{(a) Contour plot of inelastic scattering intensity
measured in Nd$_2$BaNiO$_5$ at $T=55$~K at the 1-D AF zone-center
$h=2.5$ in a series of constant-$Q$ scans (20 counts/3 min.
contour steps). The symbols show the energies of Haldane
(circles), and two crystal-field excitations (triangles), as
determined from fitting a model cross section to individual scans.
(b) The result of a global fit to the data shown in (a).}
\label{tranmap1}
\end{figure}

\begin{figure}
\caption{ (a) $k$-dependence of excitation intensities measured in
Nd$_2$BaNiOI$_5$ at $T=55$~K, as deduced from model cross section fits to
individual const-$Q$ scans (see Fig.~\protect\ref{tranmap1}a). (b)
$k$-dependence of Haldane gap energy, determined in the same fashion. Solid
lines are guides for the eye.}\label{results1}
\end{figure}

\begin{figure}
\caption{(a) Contour plot of inelastic scattering intensity
measured in Nd$_2$BaNiO$_5$ at $T=35$~K at the 1-D AF zone-center
$h=1.5$ in a series of constant-$Q$ scans (20 counts/3 min.
contour steps). The symbols show the energies of Haldane
(circles), and two crystal-field excitations (triangles), as
determined from fitting a model cross section to individual scans.
(b) The result of a global fit to the data shown in (a).}
\label{tranmap2}
\end{figure}

\begin{figure}
\caption{ (a) $k$-dependence of excitation intensities measured in
Nd$_2$BaNiO$_5$ at $T=35$~K, as deduced from model cross section
fits to individual const-$Q$ scans (see
Fig.~\protect\ref{tranmap2}a). (b) $k$-dependence of Haldane gap
energy, determined in the same fashion. Solid lines are guides for
the eye.}\label{results2}
\end{figure}

\begin{figure}
\caption{Measuring the transverse modulation of the Haldane-gap excitations
intensity: (a) Constant-$E$ scan measured in Nd$_2$BaNiO$_5$ at $T=55$~K near
the 1-D AF zone-center $h=1.5$ at $\hbar \omega=11.5$~meV (open circles). The
background is measured off the 1-D AF zone-center (solid circles). (b)
Transimission curve for the scan shown in panel (a), measured as explained in
the text.}\label{exabs1}
\end{figure}

\begin{figure}
\caption{Transverse modulation of the Haldane-gap excitations intensity:
Constant-$E$ scans measured in Nd$_2$BaNiO$_5$ at $T=55$~K near the 1-D AF
zone-centers $h=1.5$ (a) and $h=2.5$ (b) at $\hbar \omega=11.5$~meV (symbnols).
The measured intensity has been corrected for absorbtion and the measured
background has been subtracted. Solid lines are fits to a model cross section,
as described in the text. }\label{haldanemodulation}
\end{figure}

\begin{figure}
\caption{Measuring the transverse modulation of the 25 meV
excitation intensity: (a) Constant-$E$ scan measured in
Nd$_2$BaNiO$_5$ at $T=55$~K near the 1-D AF zone-center $h=2.5$ at
$\hbar \omega=25$~meV. (b) Transmission curve for the scan shown
in panel (a), measured as explained in the text.}\label{exabs2}
\end{figure}

\begin{figure}
\caption{Transverse  modulation of the 25 meV excitation
intensity: Constant-$E$ scans measured in Nd$_2$BaNiO$_5$ at
$T=55$~K away from (a) and close to (b) the 1-D AF zone-center
$h=2.5$ at $\hbar \omega=25$~meV (symbols). The measured intensity
has been corrected for absorbtion and the measured background has
been subtracted. Solid lines are fits to a model cross section, as
described in the text. }\label{cfmodulation}
\end{figure}

\begin{figure}
\caption{Logarithmic contour plot of inelastic neutron scattering
intensity measured in Nd$_2$BaNiO$_5$ at $T=55$~K near the 1-D AF
zone-center $h=1.5$. Circles and squares show peak positions as
they appear in constant-$E$ and constant-$Q$ scans, respectively.
The bars are peak widths at half height. Solids lines are guides
for the eye. Dazhed lines represent the dispersion relation
expected for the Haldane excitations (parabola) and 25~meV CF mode
(straight line) in the absense of mode coupling. The dash-dot line
shows the position of what seems to be a weak, previously
unobserved CF excitation.}\label{longmap}
\end{figure}

\begin{table}
 \caption{Parameters characterizing the transverse dispersion and intensity modulation of the
 Haldane-gap and Crystal-field excitations in \nd, as obtained from global fits of a model cross section
 to series of constant-$Q$ scans.}
 \begin{tabular}{ccc}
 & $T=55$~K ${\bbox Q}=(2.5,k,0)$ & $T=35$~K ${\bbox Q}=(1.52,k,0)$\\
 \tableline
 $\Delta$ (meV)& 10.97 (0.09) & 13.26 (0.07)\\
 $A_{\rm H}$~(meV) & -0.17 (0.01) & -0.12 (0.02) \\
 $B_{\rm H}$  & -0.19 (0.14) & 0.64 (0.15) \\
 $S_{\rm H}^{(0)}$ (arb. u.)  & 1.17 (0.03) & 0.66 (0.02)\\
 \tableline
 $E_1$~(meV) & 18.33 (0.09) & 19.7 (0.1) \\
 $A_1$~(meV) & 0 (fixed) & 0 (fixed) \\
 $B_1$       & 0 (fixed) & 0 (fixed) \\
 $S_1^{(0)}$ (arb. u.)   & 0.20 (0.01) & 0.22 (0.02)\\
 \tableline
 $E_2$~(meV) & 23.8 (0.06) & 24.6 (0.09) \\
 $A_2$~(meV) & 0 (fixed) & 0 (fixed) \\
 $B_2$       & 0.72 (0.06) & 0.37 (0.05) \\
 $S_2^{(0)}$ (arb. u.)  & 0.34 (0.01) & 0.30 (0.01)\\
 \end{tabular}
 \label{results}
 \end{table}

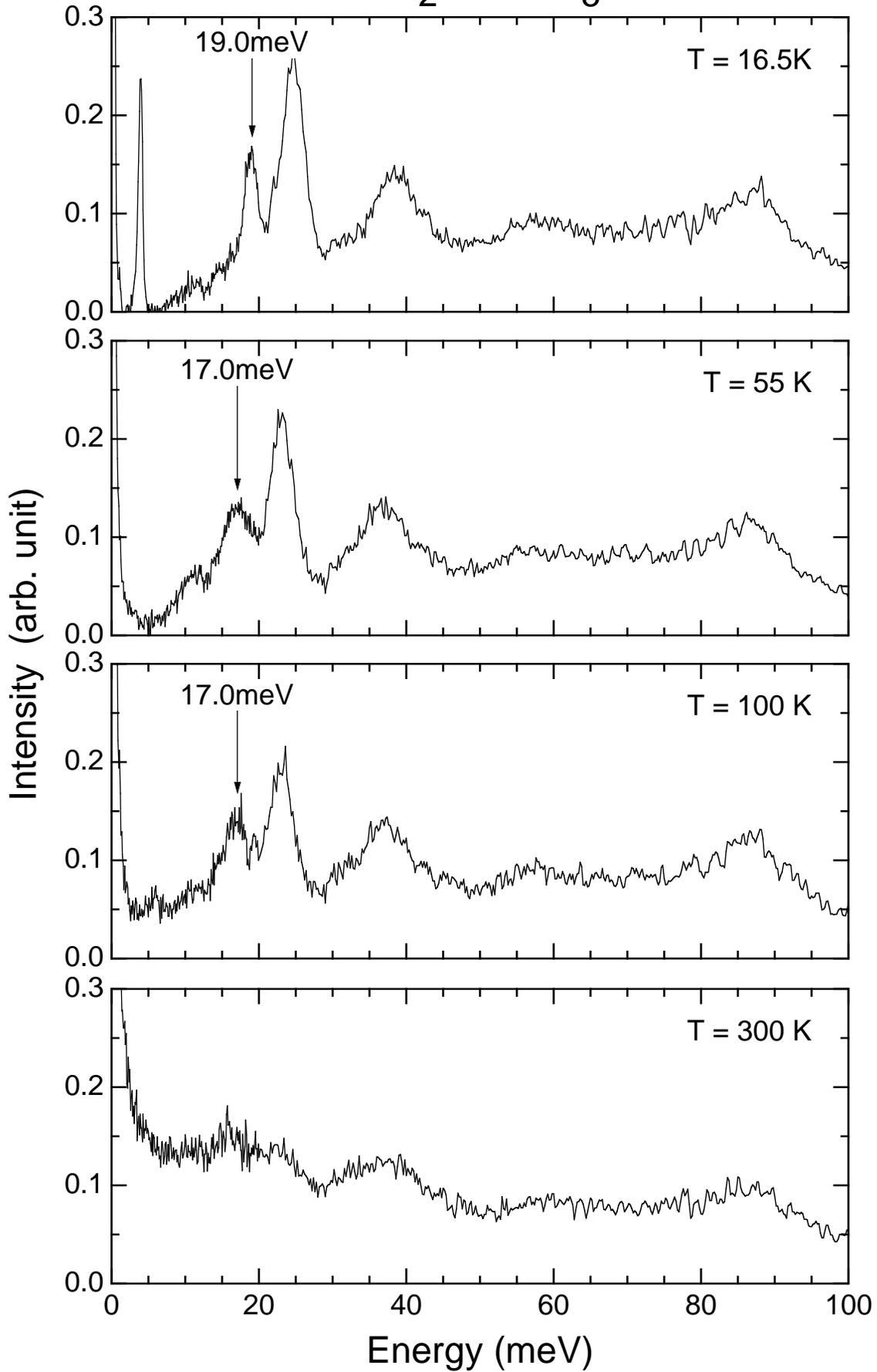
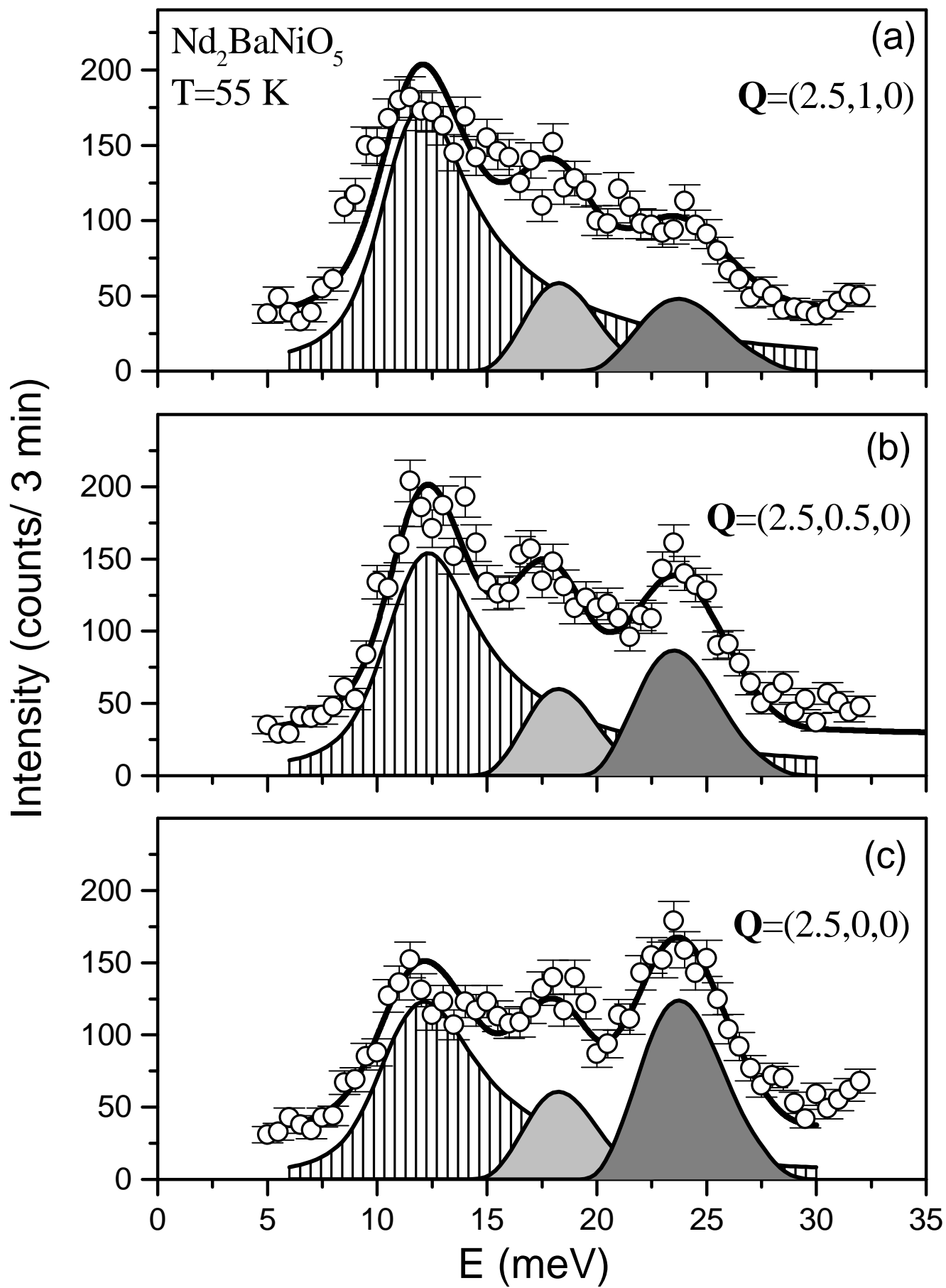
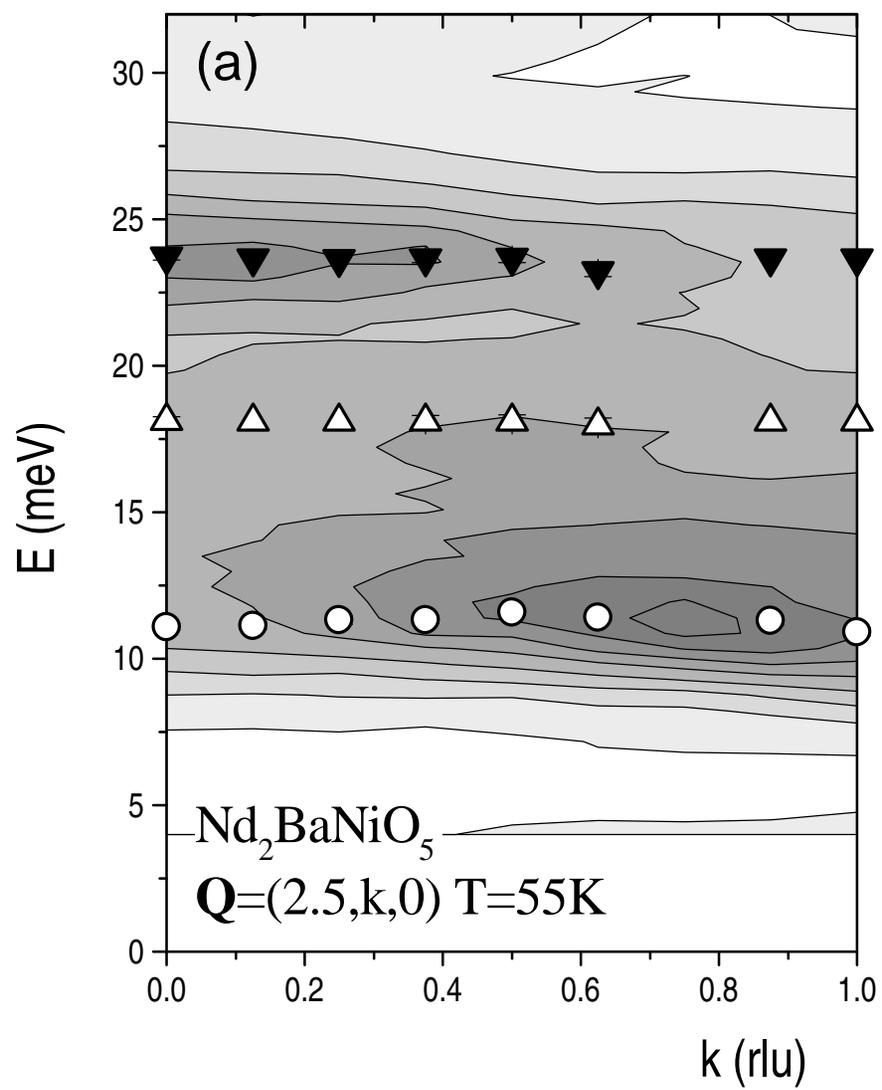
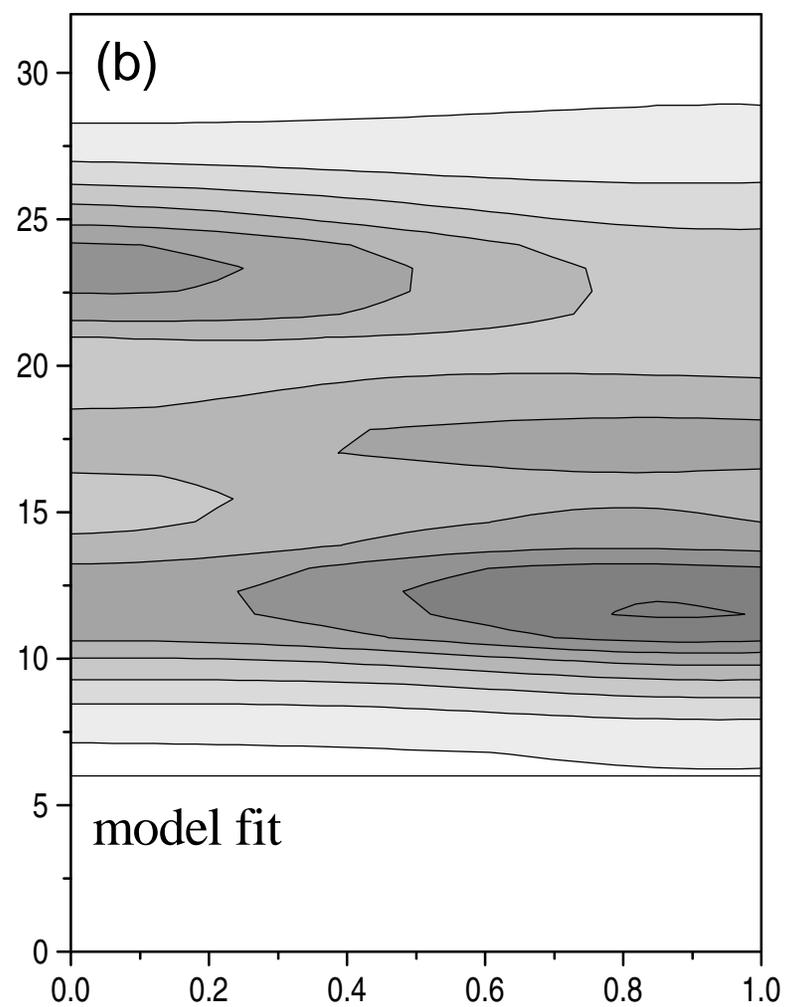
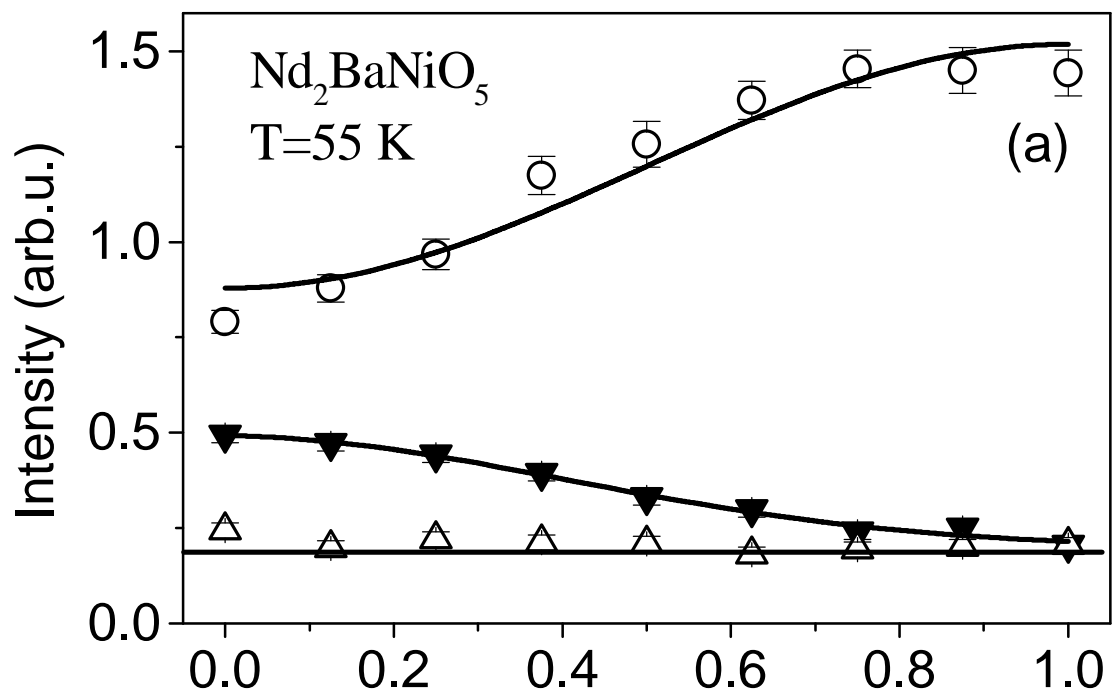
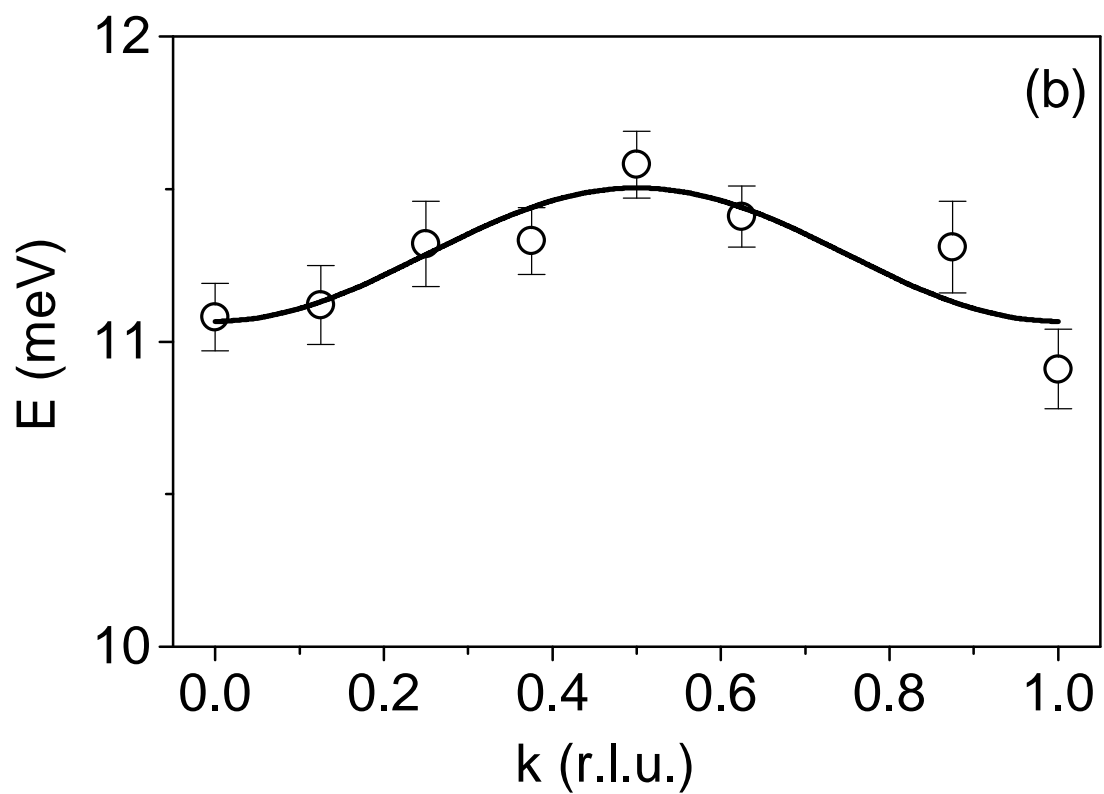
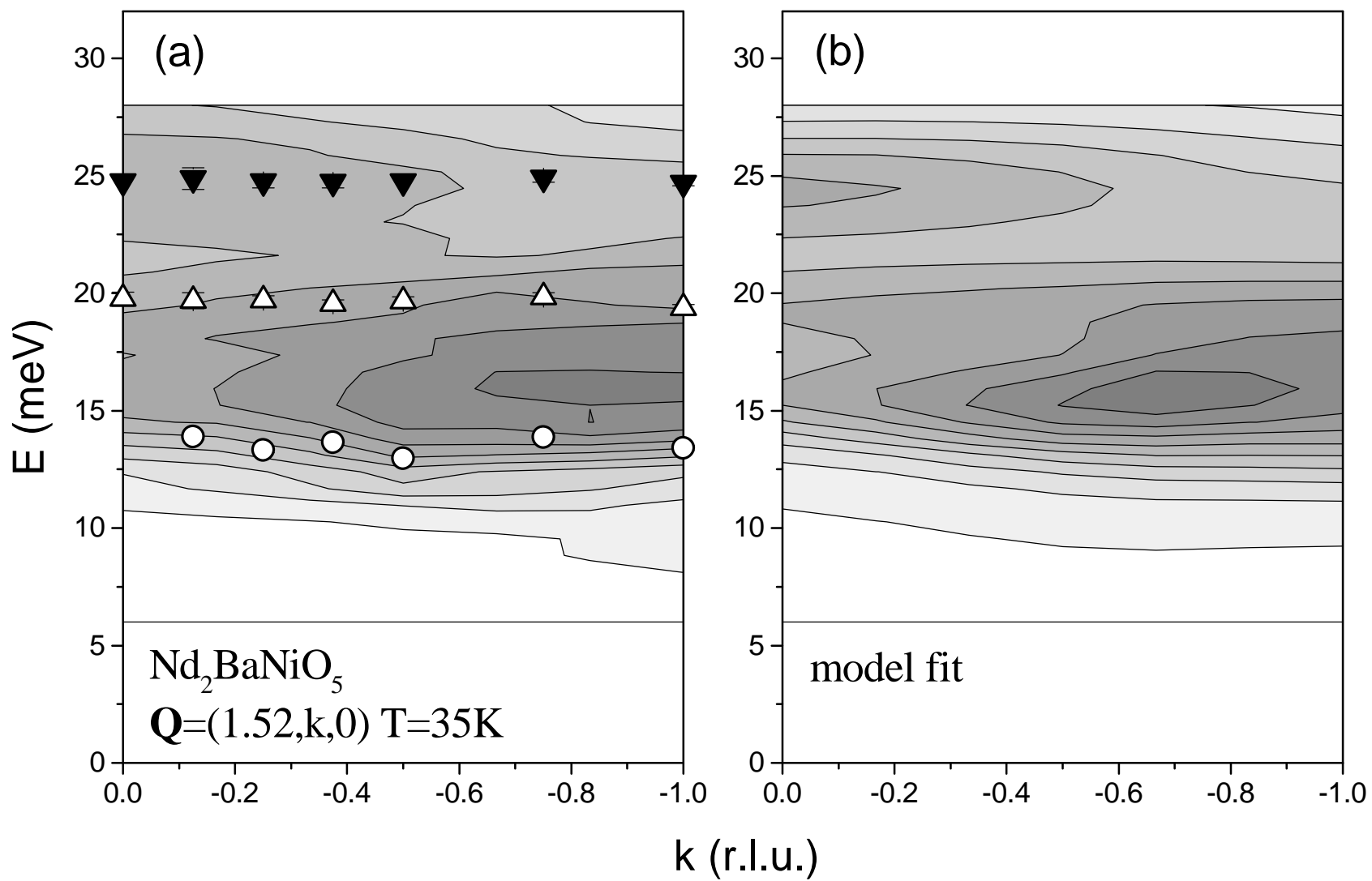
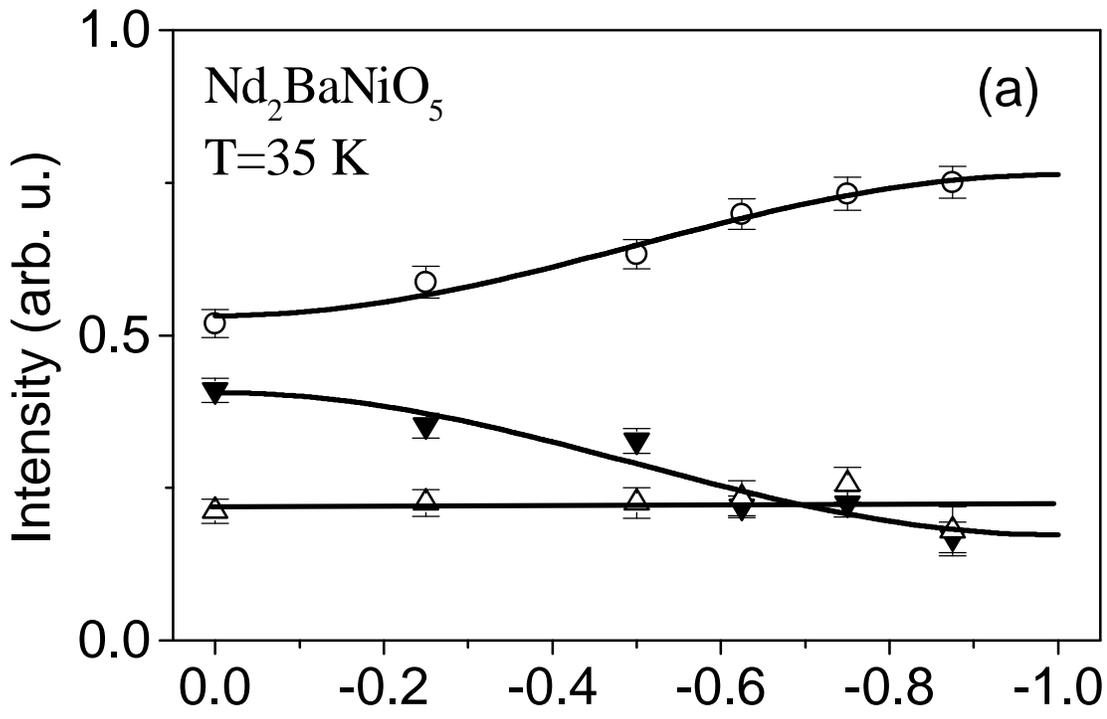
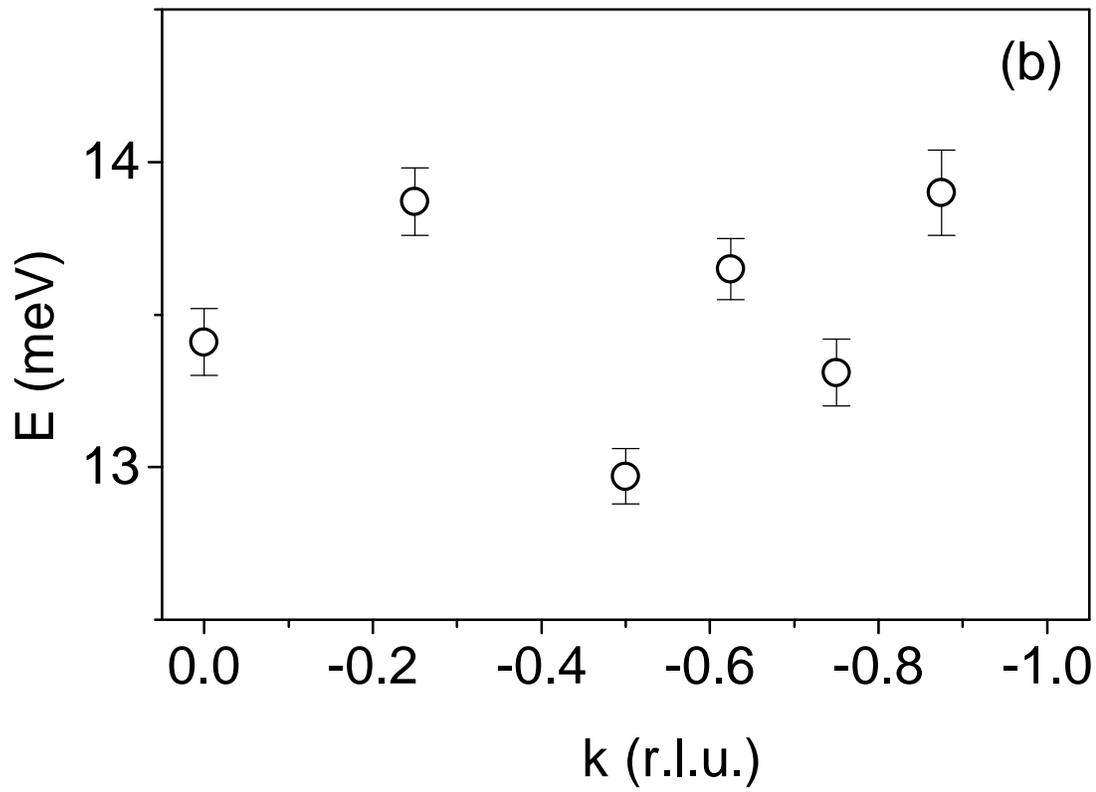
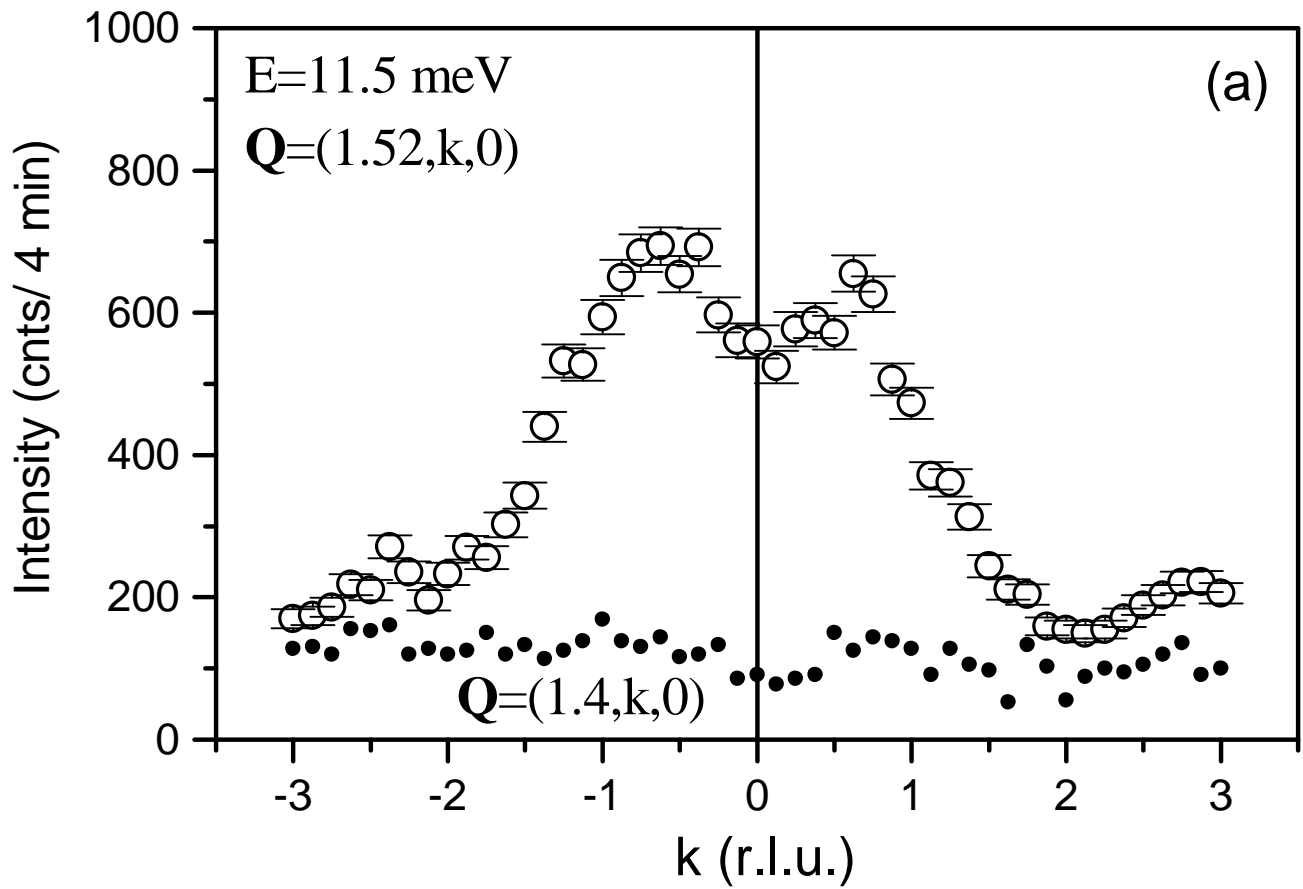
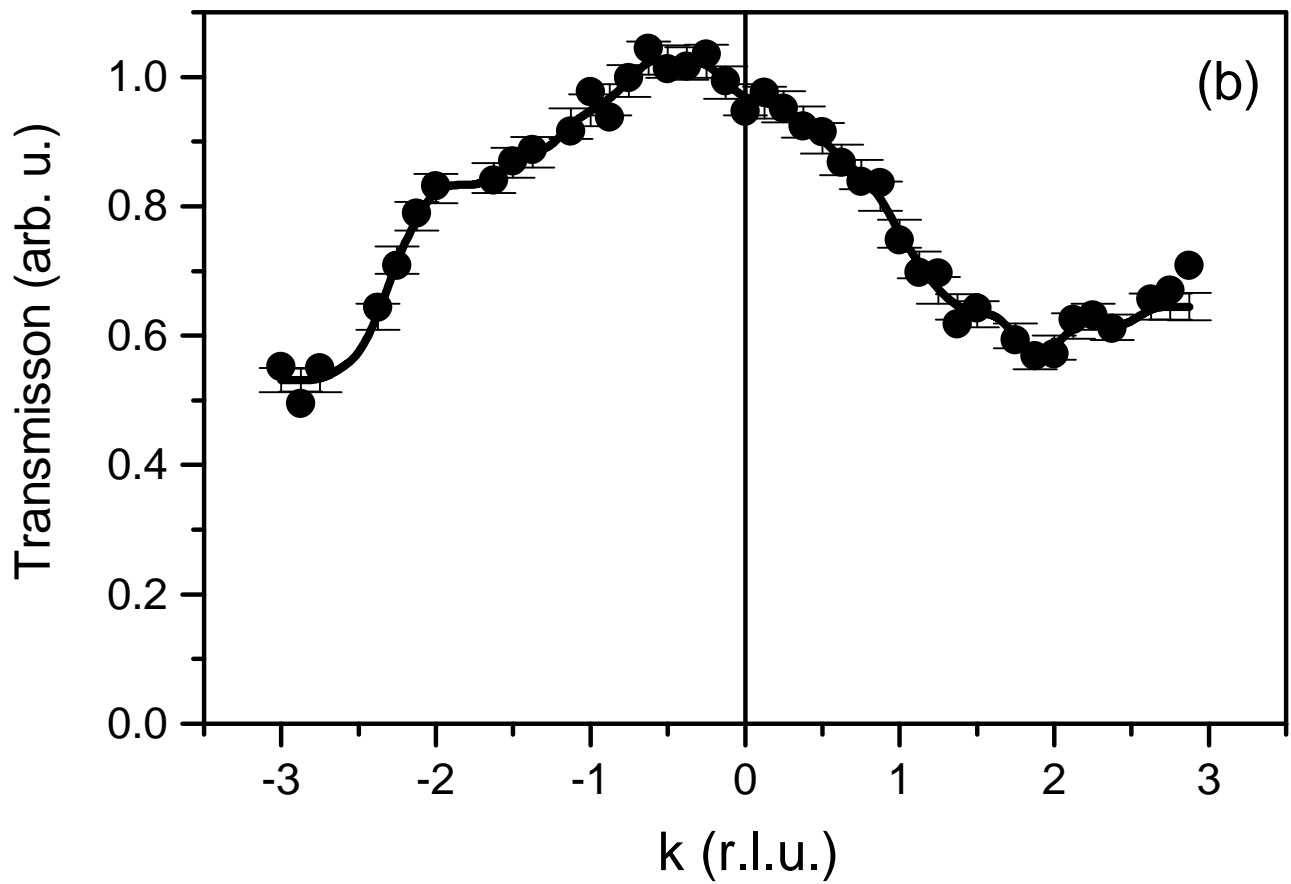
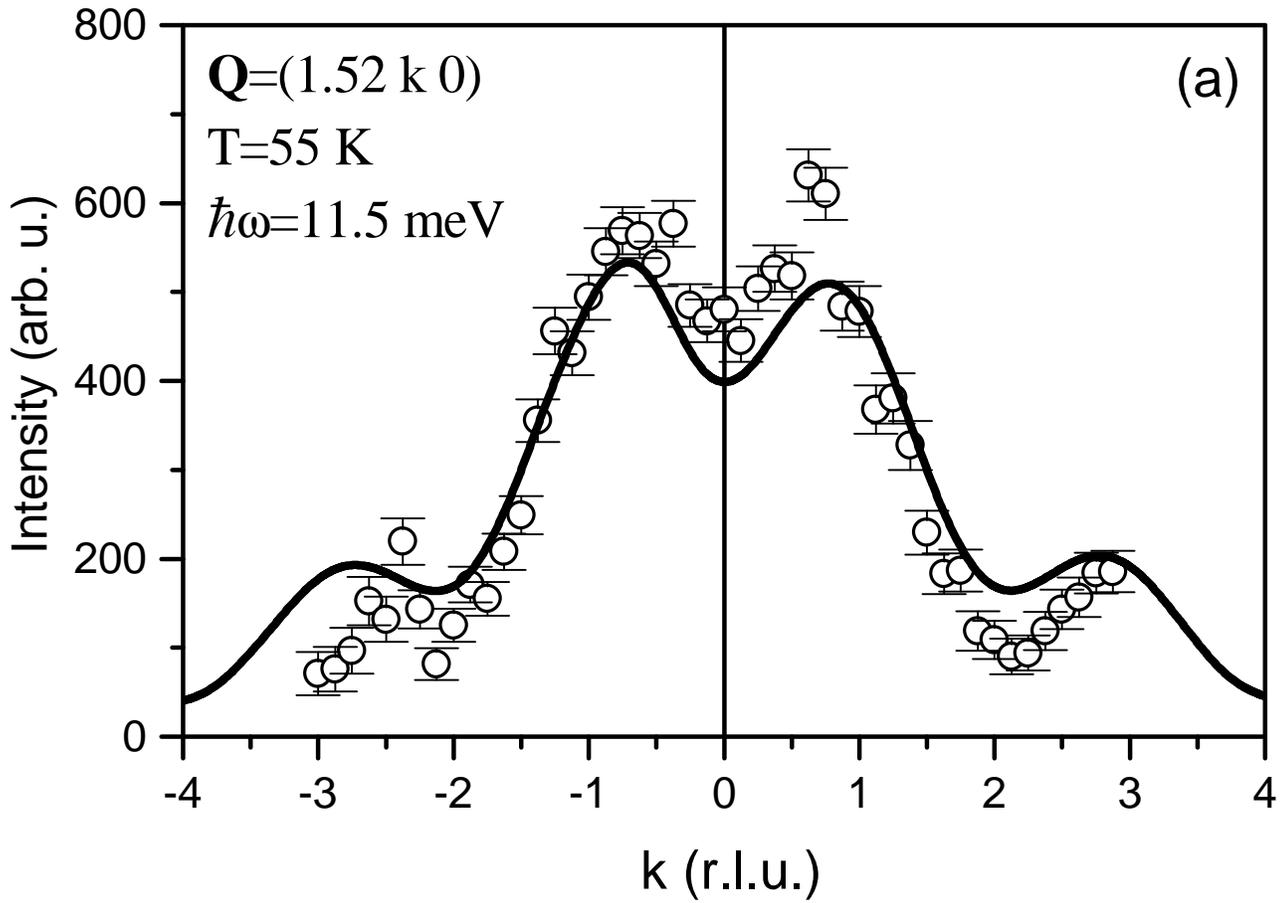
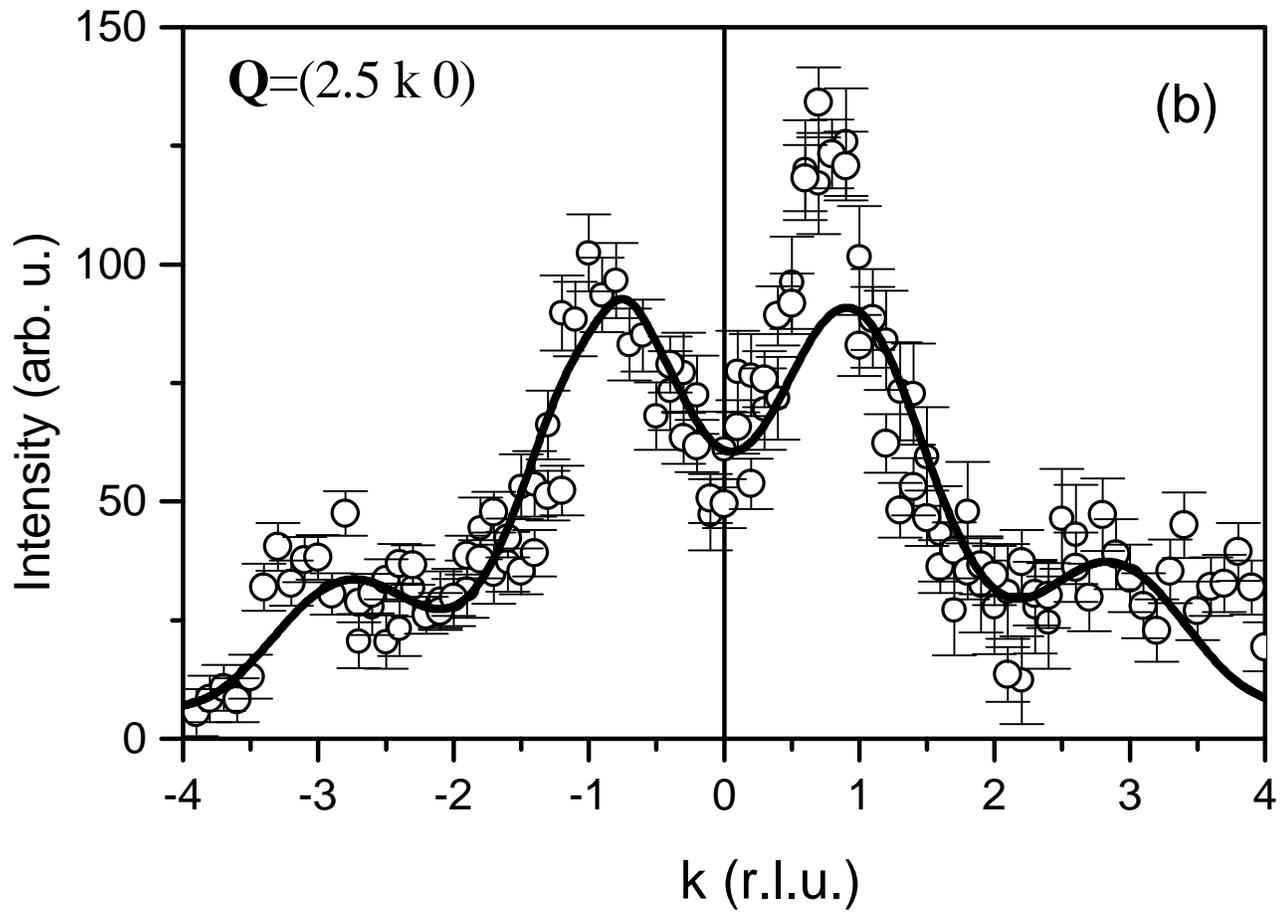
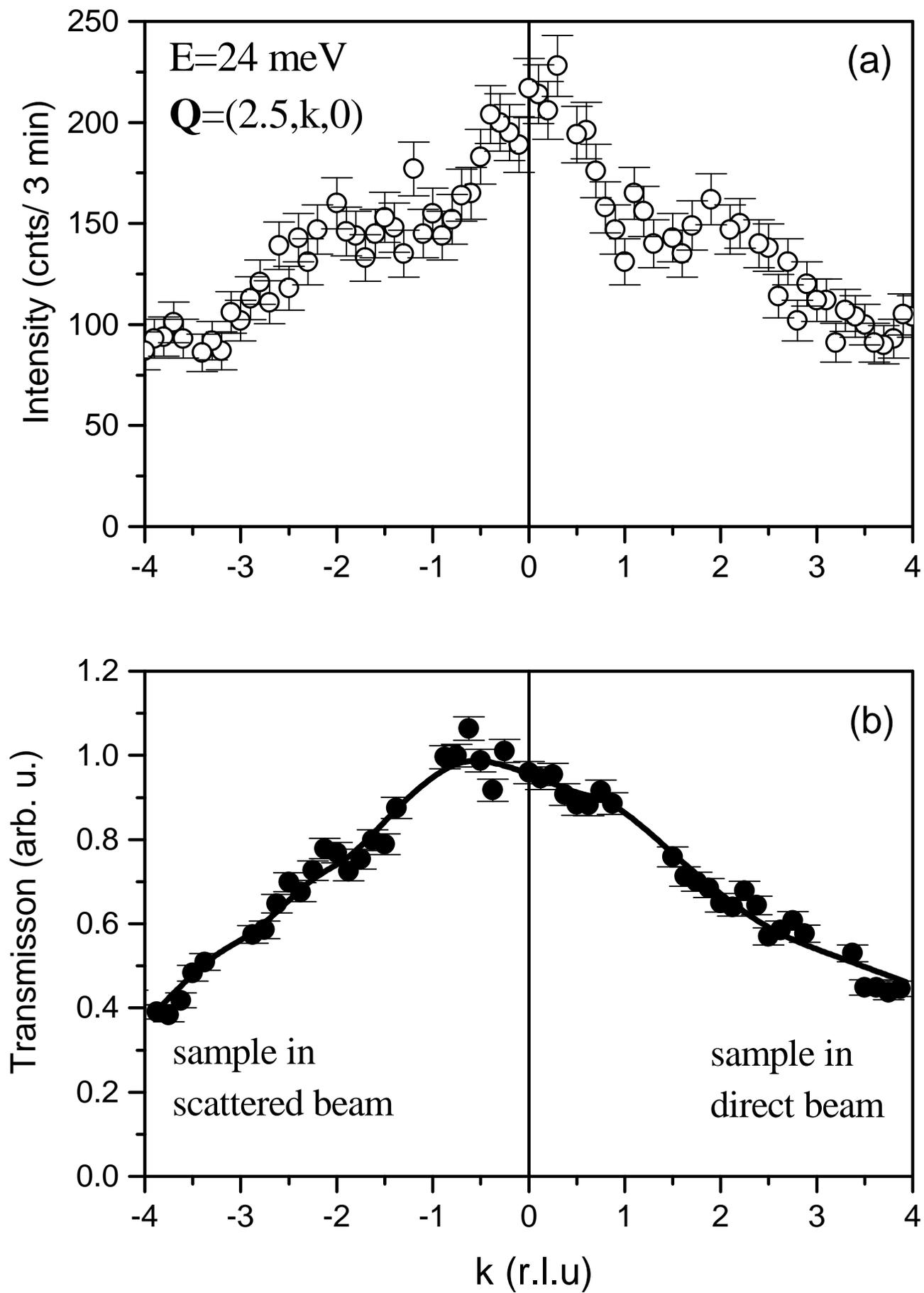
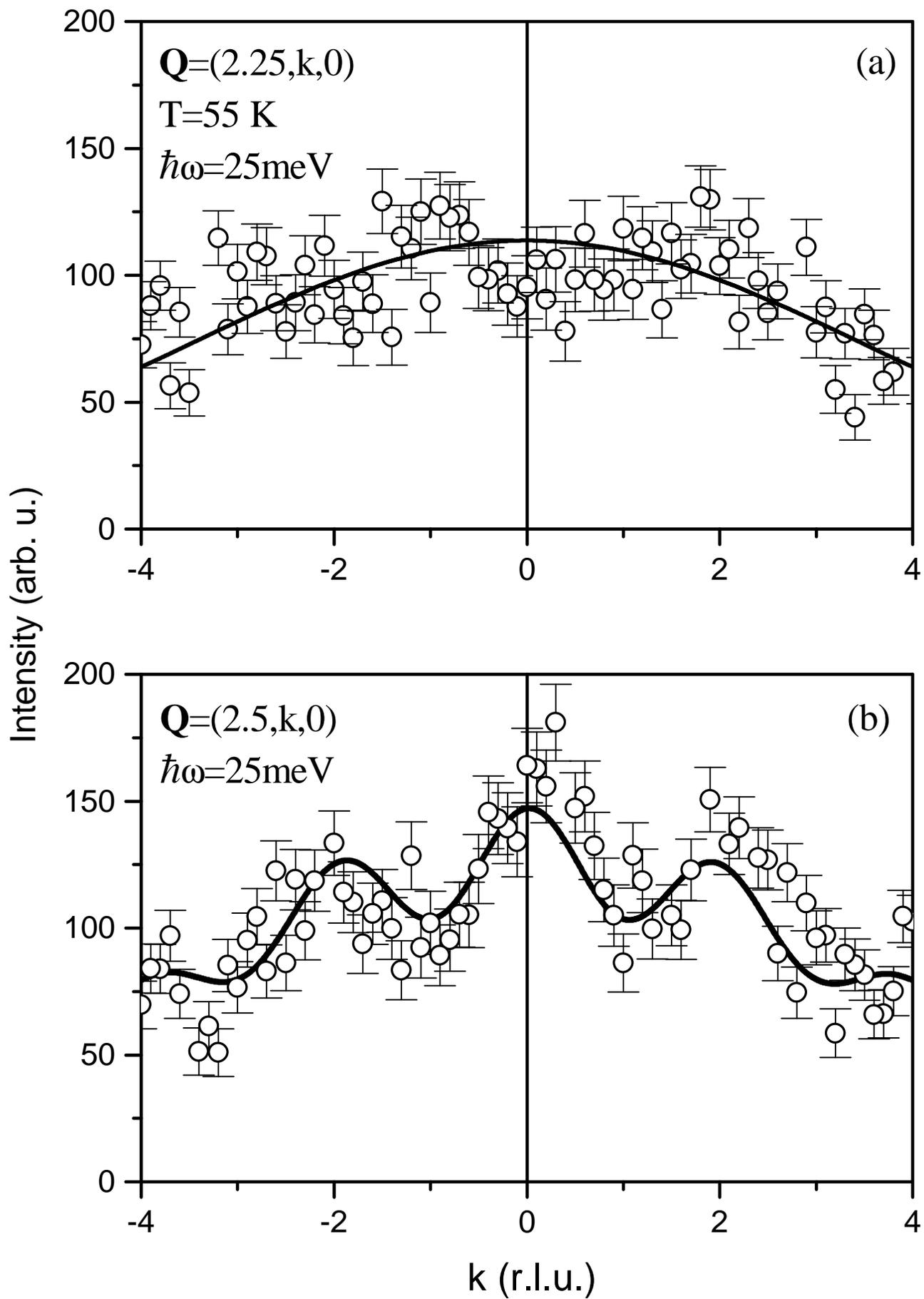
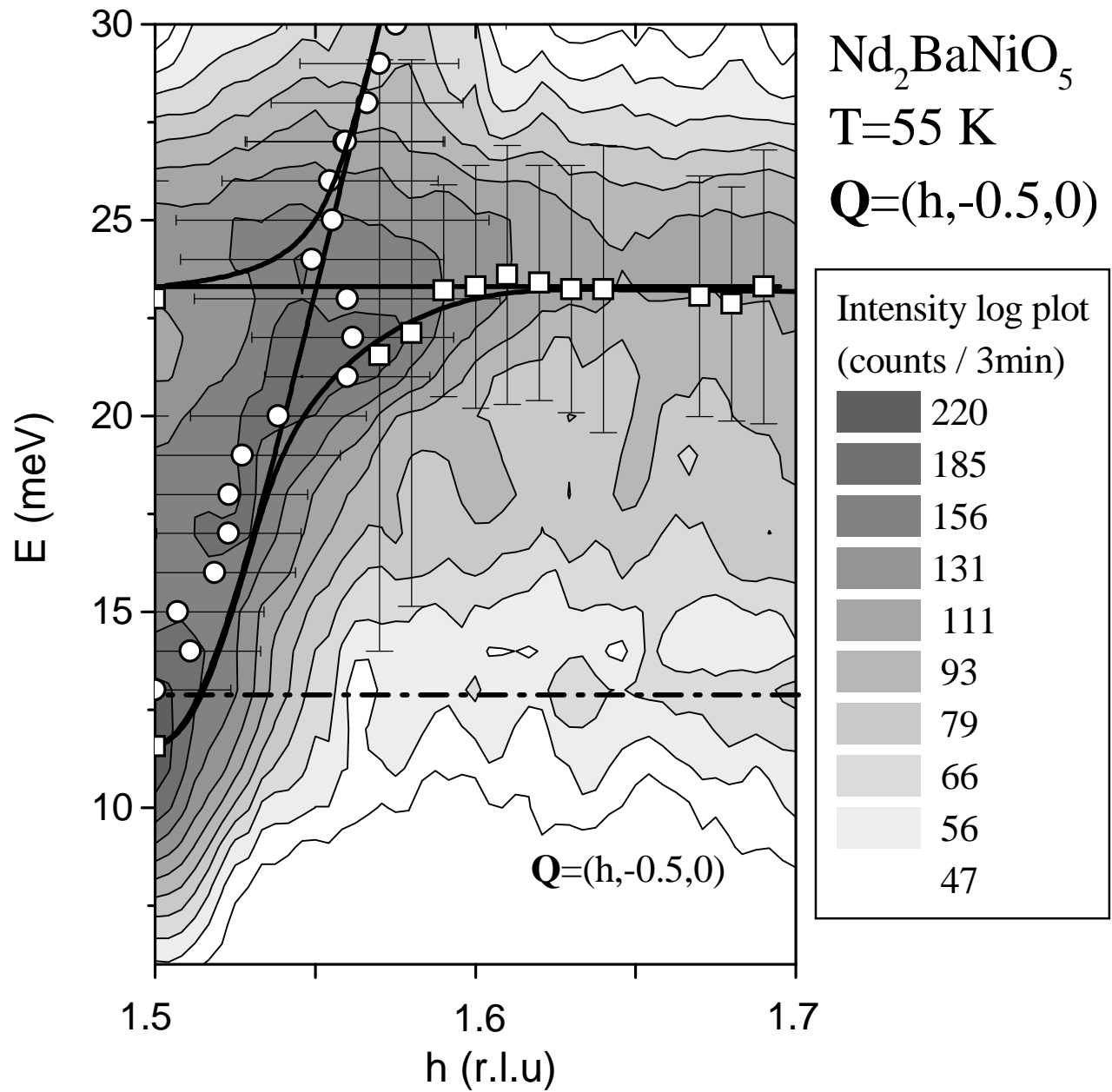
\end{document}